\documentclass{JHEP3}
\usepackage{latexsym}
\usepackage{epsfig}
\usepackage{epic}

\newcommand\cyr{%
\renewcommand\rmdefault{wncyr}%
\renewcommand\sfdefault{wncyss}%
\renewcommand\encodingdefault{OT2}%
\normalfont \selectfont} \DeclareTextFontCommand{\textcyr}{\cyr}

\title{Level Truncation and Rolling the Tachyon in the Lightcone Basis for Open String Field Theory}

\author{Theodore G. Erler\\ University of California, Santa Barbara\\ Santa Barbara,
CA 93106, U.S.A\\ E-mail:\email{terler@physics.ucsb.edu}}

\abstract{A recent paper by Gross and Erler (hep-th/0406199)
showed that by making a certain well-defined, unitary
transformation on the mode basis for the open bosonic string---one
that identifies the lightcone component of position with the
string midpoint---it is possible to render the action for cubic
string field theory local in lightcone time. In this basis, then,
cubic string field theory possesses a well-defined initial value
formulation and a conserved Hamiltonian. With this new
understanding it seems natural to study time dependent solutions
representing the the decay of an unstable D-branes. In this paper
we study such solutions using level truncation of mode oscillators
in the lightcone basis, finding both homogenous solutions by
perturbatively expanding the string field in modes $e^{nt}$, and
inhomogenous solutions by integrating the equations of motion on a
lattice. Truncating the theory to level $(\tilde{2},\tilde{4})$ in
$\alpha^+$ oscillators, we find time dependent solutions whose
behavior seems to converge to that of earlier solutions
constructed in the center of mass basis, where the cubic action
contains an infinite number of time derivatives. We further
construct time-dependent inhomogeneous solutions including all
fields up to level $(\tilde{2},\tilde{4})$. These solutions at the outset display
rather erratic behavior due to an unphysical instability introduced by truncating the theory at the linear level. However upon truncating away the field responsible for the instability, we find more reasonable solutions which may possibly represent an approximation to tachyon matter. We conclude with some discussion of future directions.}


\begin{document}
\def\a{\tilde{\alpha}}
\def\M{\mathrm{M}}
\def\P{\mathrm{P}}
\def\k{\kappa}
\def\EQN#1{eq.(\ref{eq:#1})}
\def\half{\!\matrix{\frac{1}{2}}}
\def\fraction#1#2{\matrix{\frac{#1}{#2}}}
\def\der#1{\frac{\partial}{\partial #1}}
\def\p2{\!\fraction{\pi}{2}\!}
\def\b{\tilde{b}}
\def\c{\tilde{c}}
\def\e{\epsilon}
\def\cm{\mathrm{cm}}

\section{Introduction}
String field theory has seen a remarkable revival following the
conjectures of Sen\cite{Sen}, which have revealed (upon extensive
numerical work) that string field theory provides a framework
where quite distinct string backgrounds can be formulated in terms
of a single set of underlying degrees of freedom. The most famous
example of this is the closed string vacuum solution of open
bosonic string field theory\cite{Witten}. This solution describes
the state of the open string after all unstable branes in the
bosonic theory are allowed to decay, and as such must represent
the vacuum of the closed bosonic string without any open string
excitations. Indeed, numerical analysis in the level truncation
scheme has both revealed the absence of open string
states\cite{Ellwood,Moeller} and verified that the difference in
energies between the unstable D25-brane vacuum and the stable
vacuum is precisely the energy density of the space filling D25-brane, within a fraction of a
percent\cite{Moeller-Taylor,Gaiotto}. For a useful review of open
bosonic string field theory and related work see
ref.\cite{Review}.

String theory must eventually be able to address cosmological
questions, and hence it seems crucial to understand the role of
time and dynamical solutions in the theory. String field theory,
incorporating some measure of background independence, seems to be
a natural framework in which to contemplate these questions. One
particularly simple problem is the nature of the time-dependent
decay of the unstable D25-brane in the bosonic theory.
Sen\cite{Rolling} proposed a remarkably simple boundary conformal
field theory describing such a process, whereby the tachyon rolls
homogenously off the unstable maximum towards the closed string
vacuum, but does not cross over in finite time. At late times, the
resulting ``tachyon matter'' describes a pressureless
gas\cite{Tachyon-Matter} whose cosmological implications have been
explored quite extensively in the literature\cite{Cosmology}.

Surprisingly, open bosonic string field theory has proven to be
quite inept at recovering this physically interesting solution.
Studies of the rolling tachyon process in string field theory have
approached the problem from many perspectives: inverse Wick
rotating a marginal solution\cite{Rolling,Zwiebach-Time},
perturbative expansion of the field equations in modes\cite{Hata},
and studies of time-dependence in $p$-adic
models\cite{Zwiebach-Time,Yang}. All of this work has revealed a
consistent picture, though one drastically different from Sen's:
the tachyon rolls off the unstable maximum, speeds quickly through
the closed string vacuum and then far up the steep side of the
potential, in fact quite a bit further than the height of the
unstable vacuum, after which a sequence of oscillations of
diverging amplitude ensues. Strangely, in this process the tachyon
can roll arbitrarily far into the negative, unbounded side of the
cubic-like potential, yet somehow feel the urge to turn around and
roll back up towards the unstable vacuum! The pressure of the
string field does not seem to vanish at late times, but rather
oscillates with diverging amplitude\cite{Yang}.

This type of time dependent behavior is at the very least strange,
if not catastrophic. It is not difficult to see where the problem
lies. As is well known, the cubic vertex describing the
interaction of the open bosonic string contains an infinite number
of derivatives in both time and space. As a result, the kinetic
energy for the truncated open string theory (as well as for
$p$-adic models) is not positive definite; this explains in
particular how the tachyon can roll up to a height even {\it
greater} than that of the unstable vacuum where it originated, yet
apparently manage to conserve energy. The type of instability
we're witnessing is in fact generic in any theory whose Lagrangian
depends nontrivially on any more than first time derivatives, as
was discovered over $1.5$ centuries ago by
Ostrogradski\cite{Ostrogradski}. To see the problem, consider a
Lagrangian which depends on coordinates, velocities, accelerations
etc. up to some order $N$, $L(q,\dot{q},...,q^{(N)})$.
Ostrogradski tells us we can find the Hamiltonian by defining
phase space variables, $$Q_n=q^{(n-1)}\ \ \ \ \ P_n=\sum_{k=n}^N
\left(-\frac{d}{dt}\right)^{k-n}\frac{\partial L}{\partial
q^{(k)}}\ \ \ \ n=1,2,...N$$ representing the $2N$ initial
conditions necessary to specify a solution to the Euler-Lagrange
equations. The Hamiltonian is then\footnote{If the theory is
nondegenerate, we can invert the relation for $P_N$ to write
$q^{(N)}=\dot{Q_N}(Q_1,...,Q_N,P_N)$},
$$H=\sum_{n=1}^{N-1}P_n Q_{n+1}+P_N\dot{Q}_N-L(Q_1,...,Q_N,
\dot{Q}_N)$$ Since $P_n$ and $Q_{n+1}$ for $n<N$ are independent
phase space variables, and nothing else in the Hamiltonian depends
on $P_n$, it is clear that the first term in this equation can be
made arbitrarily negative and the Hamiltonian is unbounded from
below. So truncated open string field theory and $p$-adic models
are probably sick theories, and the erratic behavior of their
solutions should be no surprise\footnote{There has been some hope
that the $N\to\infty$ higher derivative limit may fail in
truncated string field theory and $p$-adic string theory, and that
requirements of analyticity may constrain the initial value
problem to the point where the theory has a more or less the usual
canonical structure and a stable Hamiltonian. This possibility
does not seem to be borne out by the analyzes of references
\cite{Gomis,Woodward}.}.

A recent paper by Gross and the author\cite{Erler} proposed an
alternative. There, we showed that by making a certain
well-defined, unitary transformation on the mode basis for the
open bosonic string---one that identifies the lightcone component
of position with the string midpoint---it is possible to render
the action for cubic string field theory local in lightcone time.
(In the usual mode basis, both space and time are identified with
the open string center of mass.) Thus, it would seem that the {\it
exact} string field theory has both a well-defined initial value
formulation and a Hamiltonian free of higher derivative
instabilities. This suggests that the unfortunate time-dependent
behavior we're witnessing may be an unphysical artifact of the
level truncation scheme, and the exact theory admits more
reasonable time-dependent solutions, perhaps even ones resembling
Sen's tachyon matter.

\begin{figure}[top]
\begin{center}a)\resizebox{2.8in}{1.8in}{\includegraphics{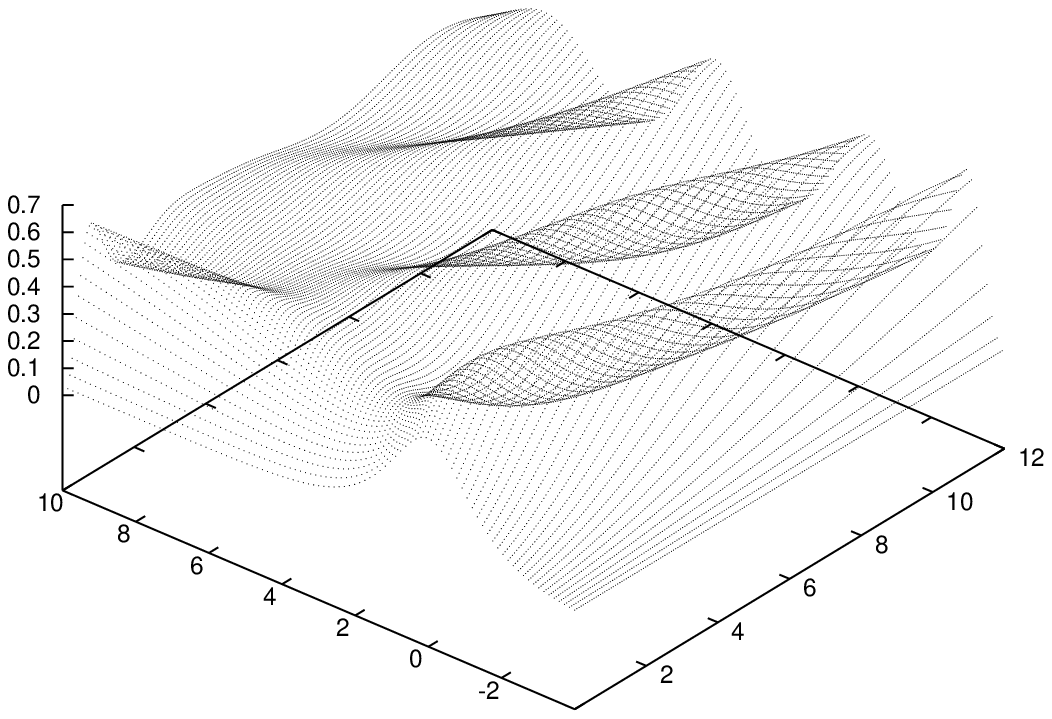}}
b)\resizebox{2.8in}{1.8in}{\includegraphics{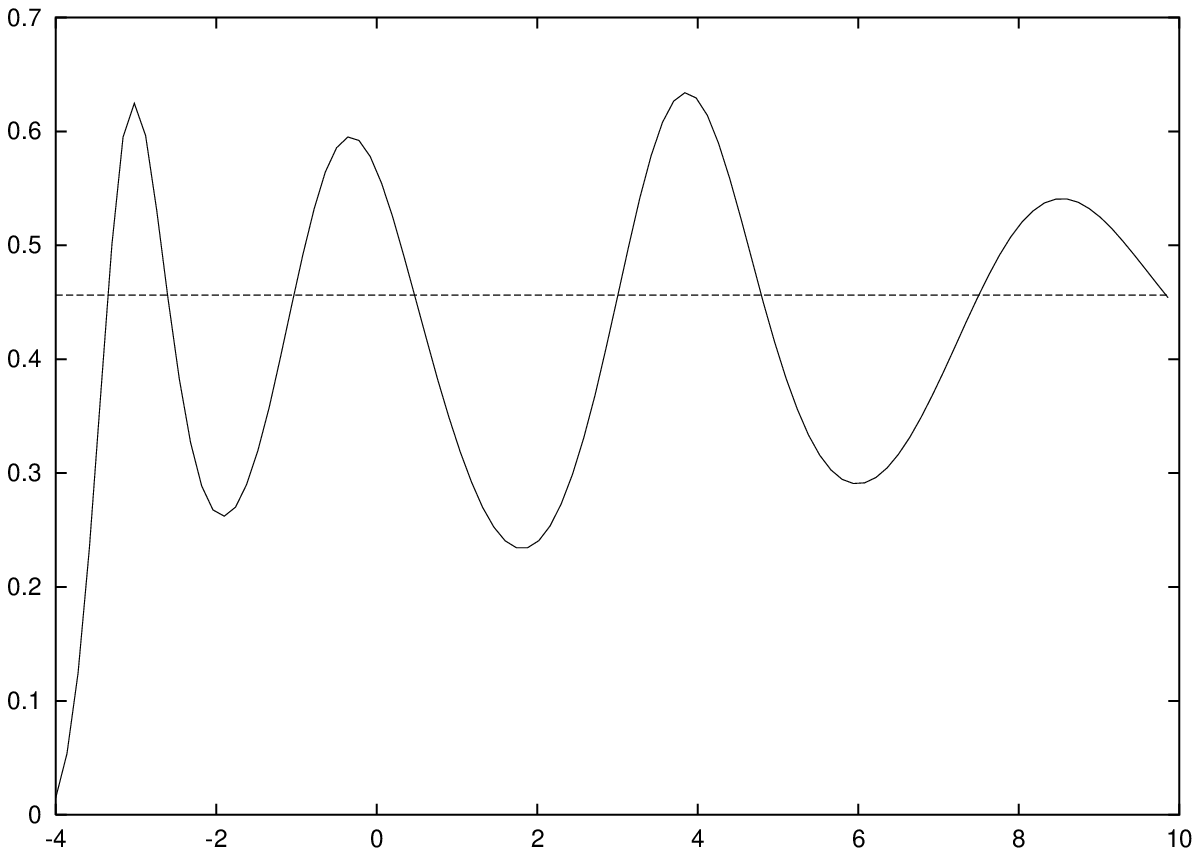}}
\end{center}
Figure 1: a) $\phi$ graphed vertically, as a function of
$(x^+,x^-)$. $x^-$ increases to the left, and ``time'' $x^+$
increases to the right. b) The profile of $\phi$ as a function of
$x^-$ at $x^+=12$. The horizontal line represents the closed
string vacuum.
\end{figure}

In this paper we investigate time-dependent solutions by
truncating string field theory in this new ``lightcone'' basis. To
get a feeling for what is involved, consider the action truncated
to level $0$ in the lightcone basis: \begin{equation} S_0=\int
dx\left[\phi(-\partial_+\partial_-+\half\partial_\perp^2-1)
\phi+\frac{2g\kappa}{3}\left(e^{-\half
V_{00}\partial_\perp^2}\phi\right)^3
\right]\label{eq:level0}\end{equation} where
$x^+=\frac{1}{\sqrt{2}}(x^0+x^1)\ x^-=\frac{1}{\sqrt{2}}(x^0-x^1)$
and $x_\perp=\{x^2,x^3,...\}$, $g$ is the open string coupling,
and $\kappa,\ V_{00}$ are numerical constants. The field $\phi$
here is a linear combination of the tachyon and lightcone
derivatives of an infinite number of massive fields. There are two
things to notice about this action. First, the Lagrangian is first
order in lightcone time derivatives $\partial_+$, which appear
only in the kinetic term. Second, the action is not Lorentz
invariant, as can be seen by the derivative dependence in the
interaction. Lorentz invariance, in fact, is only preserved in the
level truncation scheme in the infinite level limit. The curious
absence of Lorentz invariance, at the deepest level, is really
connected to the fact that string theory does not admit a
formulation in terms of spacetime fields with completely local
interactions\cite{Erler}; we can only achieve locality along a
single null direction. Thus, the loss of Lorentz invariance at
finite level is the price we have to pay for having a formulation
of string field theory which is not at the outset rendered
pathological by higher derivative instabilities and the absence of
a well-posed initial value problem.

We can easily imagine finding a dynamical solution to
eq.\ref{eq:level0} representing the decay of a D-25 brane. In
fact, looking for solutions independent of $x_\perp$,
eq.\ref{eq:level0} reduces to that of an ordinary $\phi^3$ theory
in two dimensions. To let the D25-brane decay, all we have to do
is set boundary conditions on the initial like-like surface
$x^+=0$ which displace $\phi$ off of the unstable maximum $\phi=0$
towards the ``closed string vacuum'' at $\phi=\frac{1}{g\kappa}$,
and then integrate the field equations. A numerical solution for
Gaussian initial conditions is plotted in figure 1. The field
simply flows off of the unstable maximum and oscillates in a
completely innocuous way around the stable vacuum. Of course, this
is simply a solution to $\phi^3$ theory, and we wouldn't expect
much uniquely stringy physics to emerge, but as a first pass it
isn't bad.

This paper is organized as follows. In section 2 we review some
features of the lightcone formalism of ref.\cite{Erler} which will
be relevant for our analysis. In section 3 we study the effect of
level truncation on the linear term in the string field theory
action. We find that when truncating the linear term we both loose
linear solutions to the exact theory and introduce some spurious
solutions associated with $\alpha^-$ excitations. These unstable
solutions result from the fact that the truncated equations of
motion ``drive themselves'' at their resonant frequency. In
section 4 we investigate time dependent solutions representing
D25-brane decay using a modified level expansion where we include
only $\alpha_{2n}^+$ excitations. The motivation for this particular truncation scheme is that in the infinite level limit the solutions should behave similarly to those constructed in the center of mass basis, which contains an infinite number of time derivatives. Indeed, in this truncation we find that the tachyon quickly passes through the closed string vacuum and then up the steep side of the potential to a height {\it greater} than that of the unstable vacuum where it
originated. We explain in detail how in this truncation it is possible to approximate solutions ``natrual'' in one basis by fields in the other. In section 5 we study D25-brane decay including all fields up to level $(\tilde{2},\tilde{4})$. We find that, due to a spurious linear instability introduced by level truncation, the solutions at this level behave somewhat erratically. However, by setting the field responsible for this instability to zero, we find much more reasonable solutions. The picture that emerges seems to indicate that, in comparison to solutions generated by including only scalars at level $\tilde{2}$, the tachyon rolls faster towards the closed string vacuum initially but then rapidly decelerates and does not become as large before turning around for the first time. We end with some concluding remarks.

\section{Review}
Let us review the basics of the lightcone basis introduced in
ref.\cite{Erler,Maeno}. A string field $|\Psi\rangle$ is an
element of the state space $\mathcal{H}_\mathrm{BCFT}$ of a
combined matter-ghost boundary conformal field theory describing
an open bosonic string living on a space-filling D-25 brane. The
usual basis for $\mathcal{H}_\mathrm{BCFT}$ is given by the mode
oscillators $\alpha^\mu_n, b_n, c_n$ acting on the vacuum
$|k\rangle$ describing the open string tachyon at momentum $k$
($\alpha^\mu_0 = p^\mu$). Consider a change of basis generated by
the unitary transformation,
\begin{equation} U = \exp\left[-p_+ \sum_{n=1}^\infty
\frac{(-1)^n}{2n}(\alpha^+_{2n} - \alpha^+_{-2n})\right].
\label{U}\end{equation} Under this change of basis the matter
oscillators and zero-modes transform as,
\begin{eqnarray}\a^-_n &\equiv& U\alpha^-_n U^{-1} = \alpha^-_n +
 \cos\fraction{n\pi}{2} p_+
\ \ \ \ \ n\neq 0 \nonumber\\
\tilde{x}^+ &\equiv& U x^\mu U^{-1}
 = x^+ i\sqrt{2} \sum_{n=1}^\infty\frac{(-1)^n}{2n}(\alpha^+_{2n}
 - \alpha^+_{-2n})=x^+(\p2). \label{eq:tilde_basis}\end{eqnarray}
The ghosts and other Lorentz components of $\alpha^\mu$ are
unaffected by this transformation. To simplify notation, we will
denote all components of the matter oscillators in this basis
using a tilde $\a^\mu$, even though only the minus component of
the even oscillators is affected. The vacuum $|k\rangle$
transforms into a state $|k\rangle'$:
\begin{equation}|k\rangle' \equiv \exp\left[k_+\sum_{n=1}^\infty
\frac{(-1)^n}{2n}\alpha^+_{-2n}\right]|k\rangle,\label{eq:vacuum1}\end{equation}
Since the transformation is unitary, this basis satisfies the
usual properties,
\begin{eqnarray}[\a^\mu_m, \a^\nu_{-n}] &=&
m\eta^{\mu\nu}\delta_{mn}\ \ \ \ \ [b_m,c_{-n}] =\delta_{mn} \nonumber \\
 \a^\mu_n|k\rangle' &=&
b_n|k\rangle' = c_n|k\rangle' = 0\ \ \ \ \ \ \ n>0\nonumber\\
p_\mu|k\rangle' &=& k_\mu|k\rangle'\ \ \ \ b_0|k\rangle' =
0\nonumber\\ (\a_n)^+ &=& \a_{-n}\ \ \ \ (b_n)^+=b_{-n}\ \ \ \
(c_n)^+=c_{-n}
\end{eqnarray} The zeroth matter Virasoro generator can be expressed,
\begin{eqnarray}L_0 &=& \half p^2 + p_+
\sum_{n=1}^\infty(-1)^n(\a_{2n}^+ +\a_{-2n}^+)+\sum_{n=1}^\infty \a_{-n}\cdot\a_{n}\nonumber\\
&=& \tilde{L}_0|_0 + \half p_\perp^2 +
p_+P^+(\p2)\label{L_0}\end{eqnarray} where $P(\p2)$ is the
momentum of the string midpoint, and $\tilde{L}_0$ is simply $L_0$
with the replacement $\alpha\rightarrow\a$, $|_0$ means we
evaluate the operator at zero momentum. The zeroth ghost Virasoro
takes the usual form, $$L_0^{gh}=\sum_{n=1}^\infty
n(b_{-n}c_n+c_{-n}b_n)$$ The BPZ inner product $\langle,\rangle$
satisfies the familiar relations:
\begin{eqnarray} \langle\a_{-m}^\mu\Psi,\Phi\rangle &=&
(-1)^{m+1}\langle
\Psi,\a_m^\mu\Phi\rangle\nonumber\\
\langle b_{-m}\Psi,\Phi\rangle &=& (-1)^{m}(-1)^{\Psi}\langle
\Psi, b_m\Phi\rangle\nonumber\\
\langle c_{-m}\Psi,\Phi\rangle &=& (-1)^{m+1}(-1)^{\Psi}\langle
\Psi, c_m\Phi\rangle
\end{eqnarray} where $(-1)^\Psi$ denotes the Grassmann parity of $\Psi$.
The two string vertex $\langle V_2|$ is as in the old basis after
the replacement of the oscillators and vacua with their tilded
counterparts.

String interactions are defined by the cubic vertex $\langle
V_3|\in \mathcal{H}_{BCFT}\otimes
\mathcal{H}_{BCFT}\otimes\mathcal{H}_{BCFT}$ which in the matter
sector takes the form,
\begin{eqnarray}\langle V_3^m| &=& \kappa\int dk^1 dk^2 dk^3
\delta(k^1+k^2+k^3)\langle +,k^1|'\langle +,k^2|'\langle
+,k^3|'\nonumber\\ &\ &\exp\left[-\half V_{00}^{AB}k_\perp^A\cdot
k_\perp^B- V_{m0}^{AB}\tilde{a}_{m}^A\cdot k_\perp^B
-V_{m0}^{AB}\tilde{a}_{m}^{-,A} k_-^B -\half V_{mn}^{AB}
\tilde{a}_m^A\cdot \tilde{a}_n^B\right]\label{our_vertex}
\end{eqnarray} where $\tilde{a}_m=\frac{\a_m}{\sqrt{m}}$,
the capital indices $A,B$ refer to the state space and range from
$1$ to $3$, and the mode number indices $m,n$ range from $1$ to
$\infty$, all repeated indices summed. Note that the exponential
has no dependence on $k_+$, meaning that the vertex contains no
derivatives with respect to $\tilde{x}^+$ and is therefore local
in lightcone time. The ghost component of the vertex in Siegel
gauge takes the usual form, which we write down for reference:
$$\langle V_3^{gh}|b_0^{(1)}b_0^{(2)}b_0^{(3)}=\langle -|\langle
-|\langle -|\exp\left[X_{mn}^{AB}c_m^A b_n^B\right]$$ Of course,
$\langle V_3|=\langle V_3^m|\langle V_3^{gh}|$. Explicit
expressions for the constants $\kappa, V$, and $X$ (the latter two
are called ``Neumann coefficients'') were calculated in
ref.\cite{Gross} and can be found for convenient reference for
example in ref.\cite{Review}.

At zero momentum, the lightcone basis is identical to the center of mass basis. Therefore, since the oscillator expression for the closed string vacuum is identical in either basis, at this level either basis would seem equally appropriate for investigating dynamics about the vacuum. 

For a real string field $|\Psi\rangle$ in Siegel gauge
$b_0|\Psi\rangle=0$, the string field theory action takes the
form,
\begin{equation}S=\langle \Psi|c_0(L_0 + L_0^{gh}-1)|\Psi\rangle
+\frac{2g}{3}\langle V_3||\Psi\rangle|\Psi\rangle|\Psi\rangle
\label{eq:action}\end{equation} The assumption of reality amounts
to the condition, $$\langle \Psi, A\rangle = \langle \Psi|A\rangle
$$ for any string field $|A\rangle$. The equations of motion
derived from eq.\ref{eq:action} are,
\begin{equation}\left[\half p_\perp^2 +
p_+P^+(\p2)+\tilde{L}_0|_0 +L_0^{gh}-1\right]|\Psi\rangle + g
b_0\langle\Psi|\langle\Psi||V_3\rangle = 0\label{eq:EOM}
\end{equation} In this paper we will always be looking for time dependent
solutions in Siegel gauge, so eq.\ref{eq:EOM} will be sufficient
for our purposes. These equations specify the dynamics for an {\it
infinite} number of local spacetime fields, so it seems impossible
to generate a numerical solution which includes all of them. So we
are forced to truncate the theory, including only a few of the
lightest mass fields and setting the remaining ones to zero by
fiat. The usual approach is to expand the equations in the mode
basis $\alpha_n,b_n,c_n$ where position $x$ corresponds to the
string center of mass, and keep fields in the first few low-lying
levels. As we've seen, the resulting approximate equations of
motion contain an infinite number of time derivatives, and the dynamical solutions behave very strangely. Instead, we
will expand the string field in the lightcone basis
$\a_n,b_n,c_n$, and keep only the first few levels.  As a
matter of notation we will use $(\tilde{n},\tilde{i})$ to denote
the truncation of the theory to include fields up to level $n$ and
interactions up to level $i$ in the lightcone basis, and reserve the notation $(n,i)$ (no tildes) for the corresponding truncation in the center of mass basis. The equations
of motion are then first order on lightcone time, and we can hope that the
solutions will be more reasonable.

One further comment: A solution to eq.\ref{eq:EOM}
does not, in itself, necessarily represent a full solution to the
equations of motion in open string field theory. Eq.\ref{eq:EOM}
must be supplemented by constraints on the initial conditions, which ensure that
dynamics only proceeds on a physical submanifold in phase space
where the Hamiltonian is positive (modulo the tachyon).  Unfortunately, the consistency of the constraints with time evolution at finite level turns out to be a somewhat difficult issue. We will discuss this more in section 5.

\section{The Linear Problem}
The first step in constructing time dependent solutions in the
lightcone basis is to understand what level truncation does to the
theory at the linear level. To facilitate discussion, let's
introduce a little notation. Consider a projection operator,
$\tilde{P}_n$, projecting onto states at level $n$ (in the
lightcone basis). It satisfies some basic properties,
\begin{eqnarray}\tilde{P}_m\a_n &=& \a_n\tilde{P}_{m+n}\ \ \ \
\tilde{P}_m b_n = b_n\tilde{P}_{m+n}\ \ \ \ \tilde{P}_m c_n =
c_n\tilde{P}_{m+n}\nonumber\\ \tilde{P}_n|k\rangle' &=&
\delta_{n0}|k\rangle'\ \ \ \
[\tilde{P}_n,\tilde{x}]=[\tilde{P}_n,p]=[\tilde{P}_n,b_0]=[\tilde{P}_n,c_0]=0\end{eqnarray}
Similarly we have $P_n$ (without a tilde) which projects onto
level $n$ in the center of mass basis. Now define,
$$\tilde{\Pi}_n=\sum_{i=-\infty}^n\tilde{P}_i$$
$\tilde{\Pi}_n$ truncates any state up to level $n$ in the
lightcone basis (or for $\Pi_n$ in the center of mass basis). Truncating
the theory up to level $(\tilde{n},\tilde{3n})$ amounts
to finding some state $|\Psi\rangle = \tilde{\Pi}_n|\Psi\rangle$
which satisfies the approximate field equations,
$$\tilde{\Pi}_n\left[(L_0 +L_0^{gh}-1)|\Psi\rangle + g
b_0\langle\Psi|\langle\Psi||V_3\rangle\right] = 0 $$ In principle,
sending $n\to\infty$ we recover an exact solution.

Let's consider level truncation in the free theory ($g=0$). In the
center of mass basis, this amounts to solving the equations,
\begin{equation}\Pi_n(L_0 +L_0^{gh}-1)|\Psi\rangle = (L_0
+L_0^{gh}-1)|\Psi\rangle = 0\label{eq:lin_EOM}\end{equation} where
above we noted that $\Pi_n$ commutes with $L_0,L_0^{gh}$. Of
course, eq.\ref{eq:lin_EOM} is just the exact equations of motion.
Apparently, level truncation at the free level is trivial in the
center of mass basis. However, this is not what happens in
the lightcone basis. In this case, the truncated linear equations
are,
\begin{equation}\left[\half p^2
+\tilde{L}_0|_0 +L_0^{gh}-1\right]|\Psi\rangle + p_+ \tilde{\Pi}_n
\sum_{n=1}^\infty(-1)^n(\a_{2n}^+ +\a_{-2n}^+)|\Psi\rangle = 0
\label{eq:trunc_lin_EOM} \end{equation} Since the operator
multiplying $|\Psi\rangle$ in the second term does not preserve
level number, $\tilde{\Pi}_n$ does not commute with it, and
eq.\ref{eq:trunc_lin_EOM} is not equivalent to the exact equations
of motion.

Since eq.\ref{eq:lin_EOM} and eq.\ref{eq:trunc_lin_EOM} are not
the same, their solutions are inequivalent. This would seem
problematic, since if we can't even get the correct time dependent
solutions at the linear level, we certainly shouldn't expect to
find reliable ones once we turn on the coupling. To see the nature
of this phenomenon, consider the first nontrivial case, level
$\tilde{2}$. There, we have the string field,
\begin{equation}|\Psi\rangle = \int dk \left[\phi(k) +
\frac{i}{\sqrt{2}}(B_+(k)\a^+_{-2}
+B_-(k)\a^-_{-2})\right]|k\rangle'\label{eq:Psi_2pm}\end{equation}
Of course, there are other fields at level $2$, but they are less
interesting since they only couple to fields at higher level.
Plugging eq.\ref{eq:Psi_2pm} into eq.\ref{eq:trunc_lin_EOM} for
$n=2$, we find the
equations,\begin{eqnarray}0&=&(-\half\partial^2-1)\phi +
\sqrt{2}\partial_+B_-\nonumber\\
0&=& (-\half\partial^2 + 1)B_+ + \sqrt{2}\partial_+\phi\nonumber\\
0&=& (-\half\partial^2 + 1)B_-
\label{eq:2pm_lin_EOM}\end{eqnarray} For comparison, the analogous
equations of motion in the center of mass basis for $\phi^\cm$
(the tachyon) and $B_+^\cm,B_-^\cm$ are,
\begin{eqnarray}0&=&(-\half\partial^2-1)\phi^\cm\nonumber\\
0&=& (-\half\partial^2 + 1)B_+^\cm \nonumber\\
0&=& (-\half\partial^2 + 1)B_-^\cm
\label{eq:2pm_cm_lin_EOM}\end{eqnarray} Indeed, these equations
are similar except for the presence of extra linear couplings.

\begin{figure}[top]\begin{center}\resizebox{2.8in}{1.8in}{\includegraphics{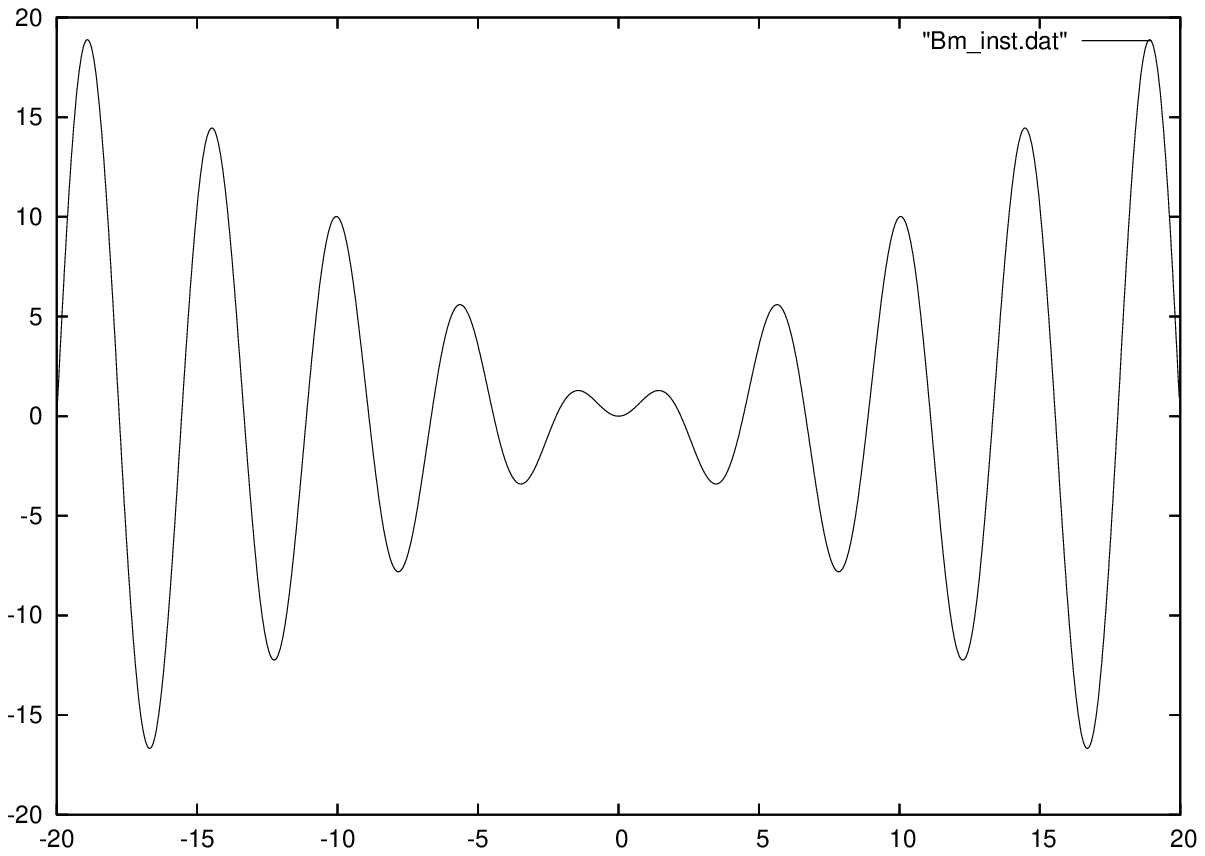}}\end{center}
Figure 2: Spurious instability in $B_+$ at level 2.
\end{figure}

So, what are the solutions of eq.\ref{eq:2pm_lin_EOM}? Consider
for simplicity spatially homogenous time dependent solutions
($t=\frac{1}{\sqrt{2}}(\tilde{x}^+ +x^-)$). Making a plane wave
ansatz for $B_-$, we can plug in and find the general solution,
\begin{eqnarray}B_-(t)&=& Ae^{i\sqrt{2}t} + Be^{-i\sqrt{2}t}\nonumber\\
\phi(t) &=& Ce^{\sqrt{2}t} + De^{-\sqrt{2}t} +
\frac{i}{\sqrt{2}}\left(Ae^{i\sqrt{2}t} - Be^{-i\sqrt{2}t}\right)\nonumber\\
B_+(t) &=& Ee^{i\sqrt{2}t} + Fe^{-i\sqrt{2}t} -
\frac{1}{\sqrt{2}}\left(Ce^{\sqrt{2}t} + De^{-\sqrt{2}t}\right)
-\frac{i}{\sqrt{2}}t\left(Ae^{i\sqrt{2}t} -
Be^{-i\sqrt{2}t}\right)\nonumber\\ \label{eq:2pm_solutions}
\end{eqnarray} where $A,B,...$ are arbitrary constants, subject to
reality. The first pair of terms in each of these equations is the
homogenous solution, corresponding to the oscillation (or decay,
as it were) of the fields at their natural frequency. The
remaining terms above come from the linear couplings in
eq.\ref{eq:2pm_lin_EOM} which can be interpreted as external
driving forces. Specifically, the field $\phi$ blows up
exponentially at late and/or early times---corresponding to the
decay of the unstable D-25 brane---but also oscillates because of
its coupling to $B_-$. $B_+$ is more interesting. In addition to
natural oscillation and tachyonic decay because of coupling to
$\phi$, we also see an awkward component where $B_+$ oscillates
with linearly diverging amplitude (see figure 2). This behavior is
directly because of $B_-$, which drives $B_+$ at its natural
frequency. The appearance of this solution is interesting, since
there does not seem to be any analogous behavior in the center of
mass basis.

Are these solutions correct? To answer this question, we have to
describe some mechanism for mapping these solutions to the center
of mass basis, and then ask whether they satisfy the exact linear
equations of motion there. A natural mapping is,
$$|\Psi\rangle=\tilde{\Pi_2}|\Psi^\cm\rangle$$ where $|\Psi^\cm\rangle$ is
the analogy of eq.\ref{eq:Psi_2pm} in the center of mass basis.
Expanding this out with the help of the transformation laws
eqs.\ref{eq:tilde_basis}-\ref{eq:vacuum1} and matching terms, we
find the relations\footnote{Note that we could have defined the
mapping to be $\Pi_2|\Psi\rangle = |\Psi^\cm\rangle$. This
turns out to give slightly different relations, differing from
eq.\ref{eq:2pm_trans} by some extra terms proportional to $B_-$.
However, the discussion goes through similarly regardless of these
definitions.},
\begin{eqnarray}\phi^\cm &=& \phi
-\frac{1}{\sqrt{2}}\partial_+B_-\nonumber\\
B_+^\cm &=& B_+ + \frac{1}{\sqrt{2}}\partial_+\phi -\half\partial_+^2 B_-\nonumber\\
B_-^\cm &=& B_-\label{eq:2pm_trans}\end{eqnarray} Plugging in our
solutions eq.\ref{eq:2pm_solutions},
\begin{eqnarray} B_-^\cm &=& Ae^{i\sqrt{2}t} +
Be^{-i\sqrt{2}t}\nonumber\\
\phi^\cm &=& Ce^{\sqrt{2}t} +
De^{-\sqrt{2}t}\nonumber\\
B_+^\cm &=& (E-\half A)e^{i\sqrt{2}t} + (F-\half B)e^{-i\sqrt{2}t}
-\frac{i}{\sqrt{2}}t\left(Ae^{i\sqrt{2}t} -
Be^{-i\sqrt{2}t}\right)\end{eqnarray} We see that $B_-^\cm$
behaves like a massive field and $\phi^\cm$ behaves like the
tachyon, as they should. However, $B_+^\cm$ still has the
diverging oscillatory part noted earlier. This piece obviously
does not satisfy the equations of motion, so it must be an
unphysical solution introduced by level truncation. The only way
to get rid of this unwanted solution is to set $B_-=0$, so the exact solution describing the natural oscillation of $B_-$ has been lost.

What happens when we include higher levels? Basically, extra
linear couplings emerge and both spurious and lost solutions
proliferate. To see what happens, it is useful to proceed more
abstractly. We make a two claims:

\noindent {\it Claim 1:} Given an exact solution to the linear
equations at level $n$ in the center of mass basis,
$\Pi_n|\Psi^\cm\rangle=|\Psi^\cm\rangle$, the field
$|\Psi\rangle=\tilde{\Pi}_n|\Psi^\cm\rangle$ satisfies
$$\tilde{\Pi}_n(L_0+L_0^{gh}-1)|\Psi\rangle = 0 $$
if an only if, $$ p_+\tilde{\Pi}_n \sum_{j=1}^n \cos\frac{\pi
j}{2}\a^+_{j}|\Psi\rangle=0$$

\noindent {\it Claim 2:} Given an solution
$\tilde{\Pi}_n|\Psi\rangle=|\Psi\rangle$ to the truncated
equations of motion in the lightcone basis,
$$\tilde{\Pi}_n(L_0+L_0^{gh}-1)|\Psi\rangle = 0 $$ the field
$|\Psi^\cm\rangle=\Pi_n|\Psi\rangle$ satisfies the exact equations
of motion if and only if, $$ p_+\Pi_n(1-\tilde{\Pi}_n)\sum_{j=1}^n
\cos\frac{\pi j}{2}\a^+_{-j}|\Psi\rangle=0$$
\bigskip

\noindent The proof of both of these statements is
straightforward. Claim 1 tells us about lost solutions in the
lightcone truncation scheme, and claim 2 tells us about spurious
solutions. The first thing to notice about these claims is that the constraints are automatically satisfied if $|\Psi\rangle$
contains no excitations proportional to $\alpha_{-2n}^-$; hence
for such states there are no spurious or lost solutions. One might
then wonder whether these constraints have any solutions with
$\alpha_{-2n}^-$ excitations at all. Surprisingly, the answer is
no. While it is fairly clear to us that this is the case, we do
not know of a demonstration simple enough to be worth presenting here since for our purposes level $\tilde{2}$ will be sufficient.

If at no level in the truncation scheme do the correct
$\alpha^-_{-2n}$ solutions emerge, in what sense can we say that
the truncated equations of motion converge to the exact ones in
the infinite level limit? The answer is that spurious solutions
are always associated with fields at the highest levels in the
particular truncation, and that by proceeding to a higher
truncation these spurious solutions disappear, only to be replaced
by new ones at even higher levels. To see how this happens,
consider level $\tilde{4}$. There, we find that $B_-$ sources the
equations of motion of a new field associated with
$\a^+_{-2}\a^-_{-2}$, and this field in turn sources $B_+$. These
new couplings conspire to cancel the unwanted diverging
oscillations we found in $B_+$ earlier. However, a new unphysical
solution is introduced at level $\tilde{4}$ by the coupling of
$B_-$, via $\phi$, to the field associated with $\a^+_{-4}$. This
unwanted solution, then, only disappears at level $\tilde{6}$ upon
introduction of $\a^+_{-4}\a^-_{-2}$. Clearly, at any level the
oscillation of $B_-$ will introduce unphysical solutions for the
fields at the highest levels, so at no stage is it consistent to
turn on $B_-$. However, it is still clear that the correct
solutions are, in some sense, emerging as the level of truncation
is increased.

While at level $\tilde{2}$ it seems okay to simply set $B_-=0$, at
higher levels fields with $\a_{-2n}^-$ excitations proliferate and
their effect on the equations of motion cannot be ignored. In this
case, it is probably not adequate to do a straightforward level
truncation. Rather, if one wanted physically correct linear
equations up to level $\tilde{n}$, one would need to include
select additional fields at higher levels, with their additional
linear couplings, so that spurious solutions at lower levels would
cancel out. At level $\tilde{2}$, for example, this could be
achieved by simply including the field associated with
$\a^+_{-2}\a^-_{-2}$.

\section{Truncation
in the $\alpha^+$ oscillators} As a first look into level truncation in the lightcone basis, it is natural to ask whether we can find time dependent solutions resembling the known (though admittedly problematic) solutions constructed in the center of mass basis, where the interaction vertex contains an infinite number of time derivatives. Indeed, it seems curious that a simple change of basis could radically alter the apparent solution space of the theory and the qualitative behavior of the dynamics. Of course, there is no real paradox here, since string field theory truncated in the center of mass basis is not equivalent to the theory truncated in the lightcone basis. However, since both truncations are an approximation to the {\it same} exact theory, truncated solutions in one basis should emerge from a suitable approximation in the other basis. It is interesting to see how this comes about, and may help us appreciate the degree to which the erratic time dependent behavior in the center of mass basis should, or should not, be taken seriously. 

As a start we investigate the rolling of the pure tachyon in the center of mass basis. The tachyon in the lightcone basis is an infinite linear combination of states excited by $\a^+_{-2n}$ oscillators:
$$|\phi^\cm\rangle=\int dk \phi^\cm(k)|k\rangle= \int
dk\phi^\cm(k) \exp\left[-k_+\sum_{n=1}^\infty
\frac{(-1)^n}{2n}\a^+_{-2n}\right]|k\rangle'$$ It is natural to suppose that if we truncate the theory in the lightcone basis in such a way that only fields corresponding to $\a^+_{-2n}$ excitations are included, we should recover solutions resembling the old solution in the center of mass basis. More precisely, the infinite level limits of $\alpha^+_{-2m}$ truncation
in the center of mass and lightcone bases should agree with each other. 

The first step is to find an appropriate time dependent solution
for the pure tachyon at level 0 in the center of mass basis, which
will serve as a standard for comparison when truncating
$\a^+_{-2n}$ excitations in the lightcone basis. Since the level 0
action for the pure tachyon contains an infinite number of time
derivatives, it does not have a well-defined initial value
formulation and it is not a priori clear how to construct generic
time dependent solutions. In the past\cite{Hata,Zwiebach-Time}, a
useful approach has been to start with a homogenous time dependent
solution in the free theory, whose equations of motion are only
second order in time derivatives, and add nonlinear
corrections to this perturbatively. The level $0$
equations of motion for the tachyon are,
\begin{equation}(-\half\partial^2-1)\phi^\cm + g\kappa e^{\half
V_{00}\partial^2}\left[e^{\half V_{00}\partial^2}
\phi^\cm\right]^2=0 \label{eq:tach_EOM}\end{equation} where $\kappa = \frac{3^{9/2}}{2^6}, V_{00}=\frac{1}{2}\ln\frac{27}{16}$ and $g$ is the open string coupling, which we will set to one. At the
linear level, an interesting homogenous time dependent solution
is,
$$\phi^\cm = \phi_1^\cm e^{\sqrt{2}t}$$ This corresponds to placing the
tachyon infinitesimally close to the unstable vacuum at
$t\to-\infty$ and then letting it roll off the unstable maximum in
the approximation that the potential is quadratic. Different
choices of $\phi_1^\cm$ corresponds to time translations of the
solution; we will set it to a convenient value in a moment. To
solve the full nonlinear equations of motion, we make a
perturbative ansatz, expanding the tachyon in modes
$e^{n\sqrt{2}t}$:
$$\phi^\cm = \sum_{n=1}^\infty \phi_n^\cm e^{n\sqrt{2} t}.$$ Plugging
this in to eq.\ref{eq:tach_EOM} implies a recursive formula for
the $\phi_n$s,
\begin{equation} \phi_n^\cm = -\frac{g\kappa e^{-\frac{1}{2}
V_{00}n^2}}{n^2-1}\sum_{k=1}^{n-1}\phi_k^\cm e^{-\frac{1}{2}
V_{00} k^2} \phi_{n-k}^\cm e^{-\frac{1}{2} V_{00}(n-k)^2}
\label{eq:tach_sol}\end{equation}

\begin{figure}[top]\begin{center}a)\resizebox{2.8in}{1.8in}{\includegraphics{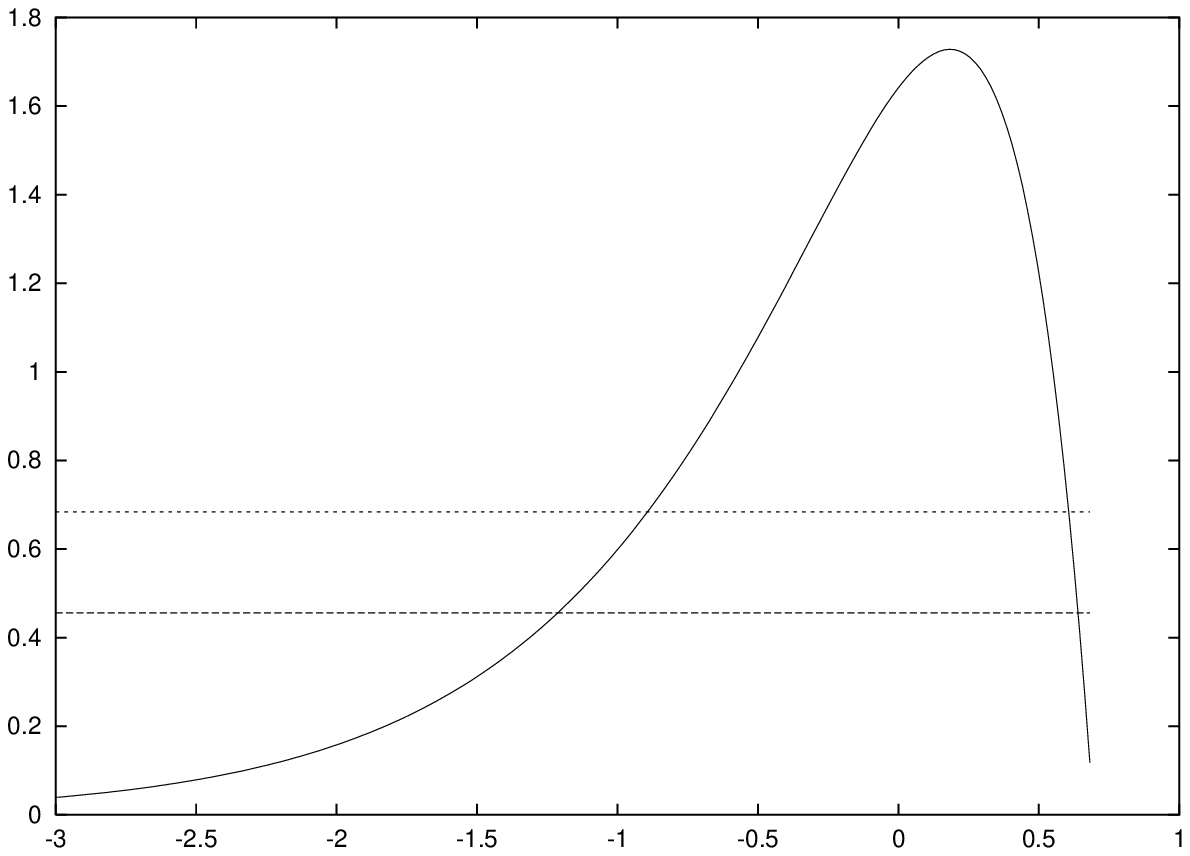}}
b)\resizebox{2.8in}{1.8in}{\includegraphics{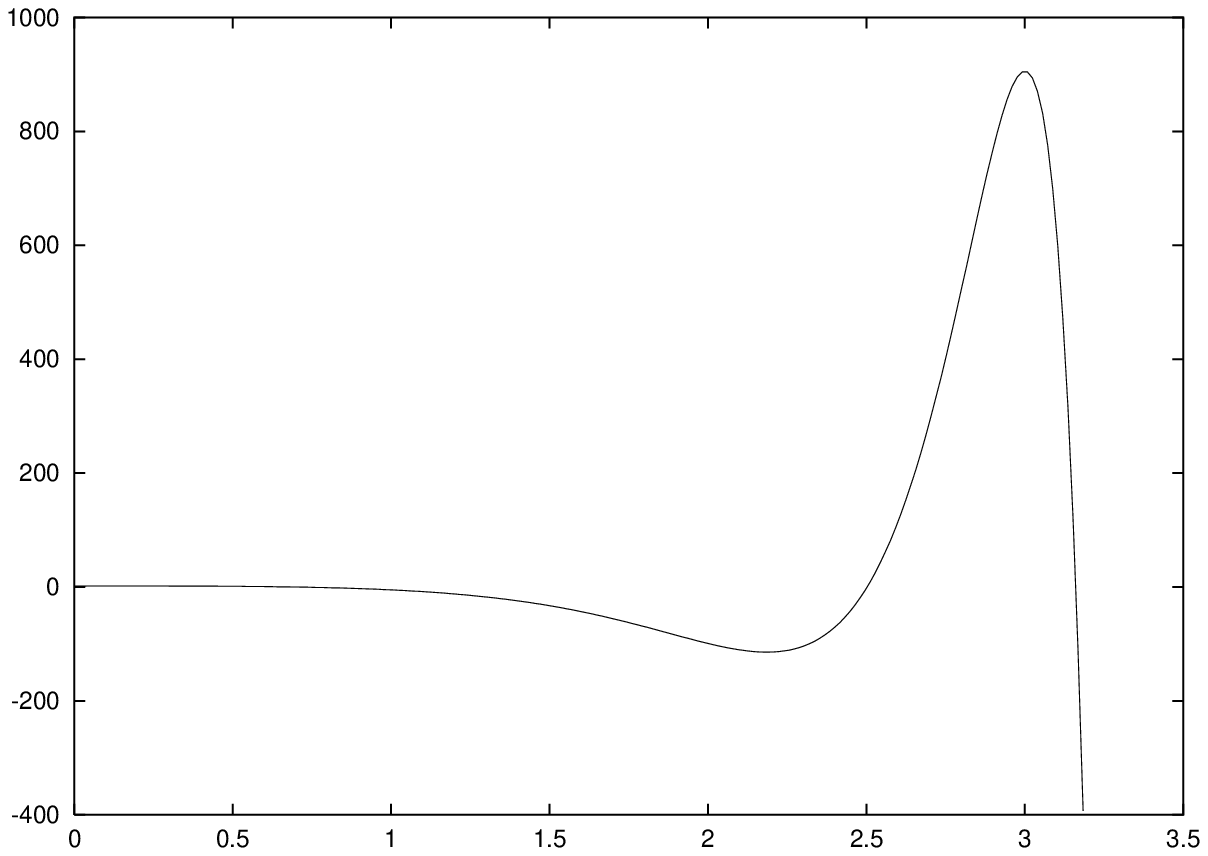}}\end{center}
Figure 3: Time evolution of the pure tachyon at level 0 in the
center of mass basis, from a) $t=-3,.6$ and b) $t=0,3.35$. These
two pictures capture the first two oscillations of the tachyon
about the closed string vacuum. The two horizontal lines in figure
3a represent the closed string vacuum and the classical turning
point in $\phi^3$ theory.
\end{figure}

\noindent The factors $e^{-\frac{1}{2}V_{00}n^2}$,
originate from the infinite number of time derivatives in the
interaction. The suppression of these factors will make the higher
$\phi_n$s much smaller than they would be in, say, $\phi^3$
theory. Since the higher modes are responsible for turning the
tachyon around after it rolls through the closed string vacuum and
the ensuing oscillations, the fact that the $\phi_n$s are small
means that the tachyon will turn around in each oscillation
comparatively late; so we would expect the tachyon to roll back
and forth about the vacuum in a sequence of diverging
oscillations. This is indeed what we find upon explicit summation
of the series as shown in figure 3. In figure 3a, we see that at the first peak of oscillation the tachyon rolls on the static potential
$V(\phi)=-\half\phi^2+\frac{g\kappa}{3}\phi^3$ to a height $64$
times the energy difference $\Delta E = .0347$ between the stable
and unstable vacua. Figure 3b shows the peak of the second oscillation of the tachyon, which reaches a staggering height of $1.6\times 10^{10}$ times
$\Delta E$! Clearly, if such time dependent
behavior was characteristic of string theory we would be in big
trouble. In ref.\cite{Hata} a similar time dependent solution for
the tachyon was constructed by expanding in $\cosh n\sqrt{2}t$
modes. This is slightly more complicated but has the advantage of
giving control over the position of the tachyon (and the velocity,
which vanishes) at $t=0$. The solution eq.\ref{eq:tach_sol},
however, will be sufficient for our purposes.

Now we turn towards constructing analogous time dependent
solutions in the lightcone basis. First consider level $\tilde{0}$, where we have only the tachyon-like field $\phi$,
$$|\Psi_0\rangle = \int dk \phi(k)|k\rangle'$$
As explained in the
introduction, the action for $\phi$ at level zero is simply that
of $\phi^3$ theory if we restrict to spatially constant solutions.
In this case, we can actually find an exact solution
for the decay process,
\begin{equation}\phi^{(\mathrm{level\ 0})}=\frac{3}{2g\kappa}\frac{1}{\cosh^2\frac{t}{\sqrt{2}}}
\label{phi_level0}\end{equation} This solution describes $\phi$
rolling off the unstable maximum, bouncing off the steep side of
the cubic potential, and then approaching the unstable maximum
again at late times. We have placed the ``bounce'' at $t=0$
corresponding to the choice, $$\phi_1 =
\phi^\cm_1=\frac{6}{g\kappa}$$ fixing the arbitrary constant
encountered earlier. 

To find a solution more closely resembling eq.\ref{eq:tach_sol} in the lightcone basis, we need to include additional fields corresponding to higher $\a^+_{-2n}$ mode excitations. Keeping $\a^+_{-2n}$ excitations up to level $\tilde{2}$ requires us to include the extra field $B_+$, which we met in eq.\ref{eq:Psi_2pm}: 
\begin{eqnarray} |\Psi_2\rangle &=& |\Psi_0\rangle +
\frac{i}{\sqrt{2}}\int dk B_+(k)\a^+_{-2}|k\rangle'\nonumber
\end{eqnarray} What are the relevant truncated equations of motion for $\phi,B_+$? The correct prescription is to write the cubic action including {\it all} fields up to level $\tilde{2}$ (including $B_-$). Then, we vary the action with respect to the {\it center of mass} fields $\phi^\cm,B_-^\cm$, and then set everything in the resulting equations of motion to zero except $\phi,B_+$. It is crucial that we vary the action with respect to the center of mass fields since ultimately we are interested in the approximate equations of motion for the pure tachyon, which is obtained by varying the action with respect to $\phi^\cm$. Thus, the relevant equations of motion are, \begin{eqnarray}0&=&\left.\frac{\delta S}{\delta\phi^\cm(x)}\right|_{\phi,B_+\neq0; B_-...=0}\nonumber\\
&=&\left.\left(\frac{\delta S}{\delta\phi(x)}+\frac{1}{\sqrt{2}}\partial_+\frac{\delta S}{\delta B_+(x)}\right)\right|_{\phi,B_+\neq0; B_-...=0}\nonumber\\
0&=&\left.\frac{\delta S}{\delta B_-^\cm(x)}\right|_{\phi,B_+\neq0; B_-...=0}=\left.\frac{\delta S}{\delta B_-(x)}\right|_{\phi,B_+\neq0; B_-...=0}\label{eq:2p_EOM1}\end{eqnarray} Here we used the chain rule and the transformation law eq.\ref{eq:2pm_trans} relating fields in the center of mass basis to the lightcone basis. Explicitly including interactions up to level $\tilde{4}$ and setting the transverse momenta to zero, these equations of motion are, 
\begin{eqnarray} 0 &=& (\partial_+\partial_- - 1)\phi+g\kappa\left[\phi^2 + G_{\phi\phi B_+}\phi\partial_-B_+ + G^1_{\phi B_+B_+}\partial^2_-B_+ B_+ +G^2_{\phi B_+ B_+}(\partial_-B_+)^2\right.\nonumber\\
&\ & \left.+\frac{1}{\sqrt{2}}\partial_+\left(G_{B_+\phi\phi}\phi\partial_-\phi + G^1_{B_+\phi B_+}\partial^2_-\phi B_+ + G^2_{B_+\phi B_+}\partial_-\phi\partial_-B_+ +G^3_{B_+\phi B_+}\phi\partial^2_-B_+\right)\right]\nonumber\\
0&=&-(\partial_+\partial_- + 1)B_+-\sqrt{2}\partial_+\phi + g\kappa G_{B_-\phi B_+}\phi B_+ \label{eq:2p_EOM}\end{eqnarray} The eight $G$s above are various powers and linear combinations of the Neumann coefficients. We can easily solve these equations by making a perturbative ansatz:
\begin{eqnarray}\phi &=& \sum_{n=1}^\infty \phi_n e^{n\sqrt{2}t}\
\ \ \ \ \ \ \
B_+ = \sum_{n=1}^\infty B_{+,n} e^{n\sqrt{2}t}\label{eq:2p_ansatz}
\end{eqnarray} where for consistency at the linear level we have,
\begin{equation}B_{+,1}=-\frac{1}{\sqrt{2}}\phi_1\end{equation}
Plugging this ansatz into eq.\ref{eq:2p_EOM} we derive recursive formulas for the various coefficients in the usual way. Summing the series, we find the tachyon profile shown in figure 4, which for comparison we have graphed alongside the level $\tilde{0}$ profile of $\phi^3$ theory and the level $0$ profile in the center of mass basis. We see that, already at level $\tilde{2}$, the rolling of the tachyon looks much more like it does at level $0$ in the center of mass basis. Strictly speaking, however, we should be comparing this result to the analogous computation including $B_+^\cm$ in the center of mass basis. However, at least in the region of convergence of eq.\ref{eq:2p_ansatz}, including $B_+^\cm$ has negligible effect on the level $0$ profile.

\begin{figure}[top]\begin{center}\resizebox{3.8in}{2.8in}{\includegraphics{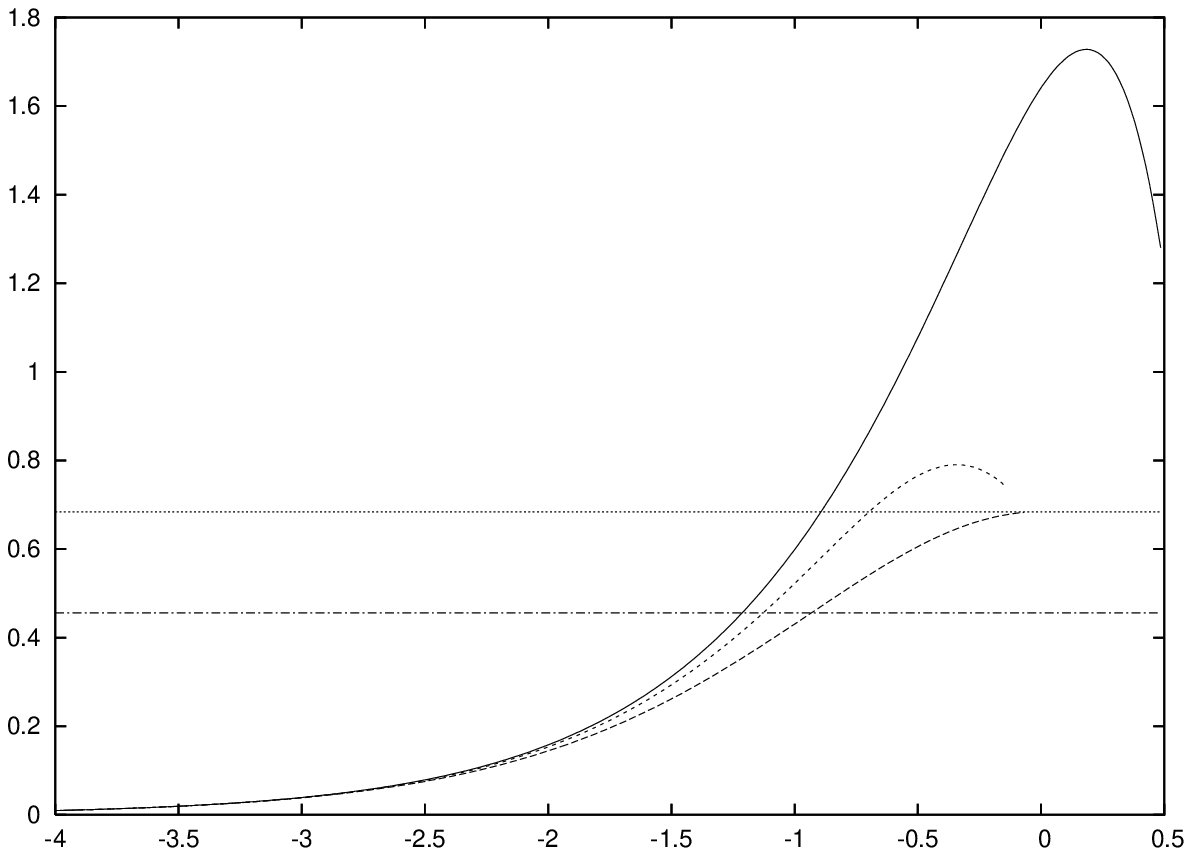}}\end{center}
Figure 4: The level $\tilde{2}$ approximation to the rolling of the pure tachyon in the center of mass basis, sandwiched between the level $\tilde{0}$ profile in the lightcone basis and the level $0$ profile in the center of mass basis. The perturbative series eq.\ref{eq:2p_ansatz} does not converge past $t\approx-.22$. The horizontal lines represent the level $0$ closed string vacuum and the turning point in the $\phi^3$ potential.\end{figure}

We emphasize that the equations of motion we have just solved,  
\begin{equation} 0=\left.\frac{\delta S}{\delta\phi^\cm(x)}\right|_{\phi^\cm,B_+^\cm\neq0; B_-^\cm...=0}=\left.\frac{\delta S}{\delta B_-^\cm(x)}\right|_{\phi^\cm,B_+^\cm\neq0; B_-^\cm...=0}=...\label{eq:EOMa}\end{equation}
are {\it not} equivalent to 
\begin{equation} 0=\left.\frac{\delta S}{\delta\phi(x)}\right|_{\phi,B_+\neq0; B_-...=0}=\left.\frac{\delta S}{\delta B_-(x)}\right|_{\phi,B_+\neq0; B_-...=0}=...\label{eq:EOMb}\end{equation} even when both are expressed in terms of fields in the same basis. The second set of equations, in particular, contain no $\partial_+$ derivatives in the interaction when written in the lightcone basis. These equations differ from each other by terms proportional the $B_-$ equation of motion---and, at higher levels, those for other $\a_{-2n}^-$ fields. If we allow the $\a_{-2n}^-$ fields to be nonzero and separately solve their equations of motion, then eq.\ref{eq:EOMa} and eq.\ref{eq:EOMb} become equivalent.

However, this suggests the following: Since the second set of equations, eq.\ref{eq:EOMb}, contain {\it no} lightcone time derivatives in the interaction, we would expect their solutions to be more nicely behaved--- resembling that of $\phi^3$ theory at level $\tilde{0}$. This suggests that it might be possible to approximate these more well-behaved solutions by truncating $\alpha^+_{-2n}$ modes in the {\it center of mass basis}. At level $2$ we must consider $\phi^\cm,B_+^\cm$. By a similar prescription to eq.\ref{eq:2p_EOM1}, the relevant equations of motion are, \begin{eqnarray}0&=&\left.\frac{\delta S}{\delta\phi(x)}\right|_{\phi^\cm,B_+^\cm\neq0; B_-^\cm...=0}\nonumber\\
&=&\left.\left(\frac{\delta S}{\delta\phi^\cm(x)}-\frac{1}{\sqrt{2}}\partial_+\frac{\delta S}{\delta B_+^\cm(x)}\right)\right|_{\phi^\cm,B_+^\cm\neq0; B_-^\cm...=0}\nonumber\\
0&=&\left.\frac{\delta S}{\delta B_-(x)}\right|_{\phi^\cm,B_+^\cm\neq0; B_-^\cm...=0}=\left.\frac{\delta S}{\delta B_-^\cm(x)}\right|_{\phi^\cm,B_+^\cm\neq 0; B_-^\cm...=0}\end{eqnarray} As usual, we solve these equations by making a perturbative ansatz. The resulting tachyon profile is shown in figure 5a, along with the profiles at level $0$, eq.\ref{eq:tach_sol}, and at level $\tilde{2}$ with $\phi,B_+$ in the lightcone basis. We can see that including $B_+^\cm$ makes the tachyon turn around well before it does at level $0$. Figure 5b shows the evolution of $\phi^\cm$ at later times. While the subsequent oscillations are still pronounced, they are much less virulent then they were at level $0$; the peak of the second oscillation reaches a potential height of only $2\times10^7\times\Delta E$, about $1000$ times smaller than the height of the second peak at level $0$. Remarkably, even computing in the center of mass basis, truncating the theory in a particular way can yield time dependent solutions whose qualitative behavior is substantially improved over the level $0$ solution.

\begin{figure}[top]\begin{center}a)\resizebox{2.8in}{1.8in}{\includegraphics{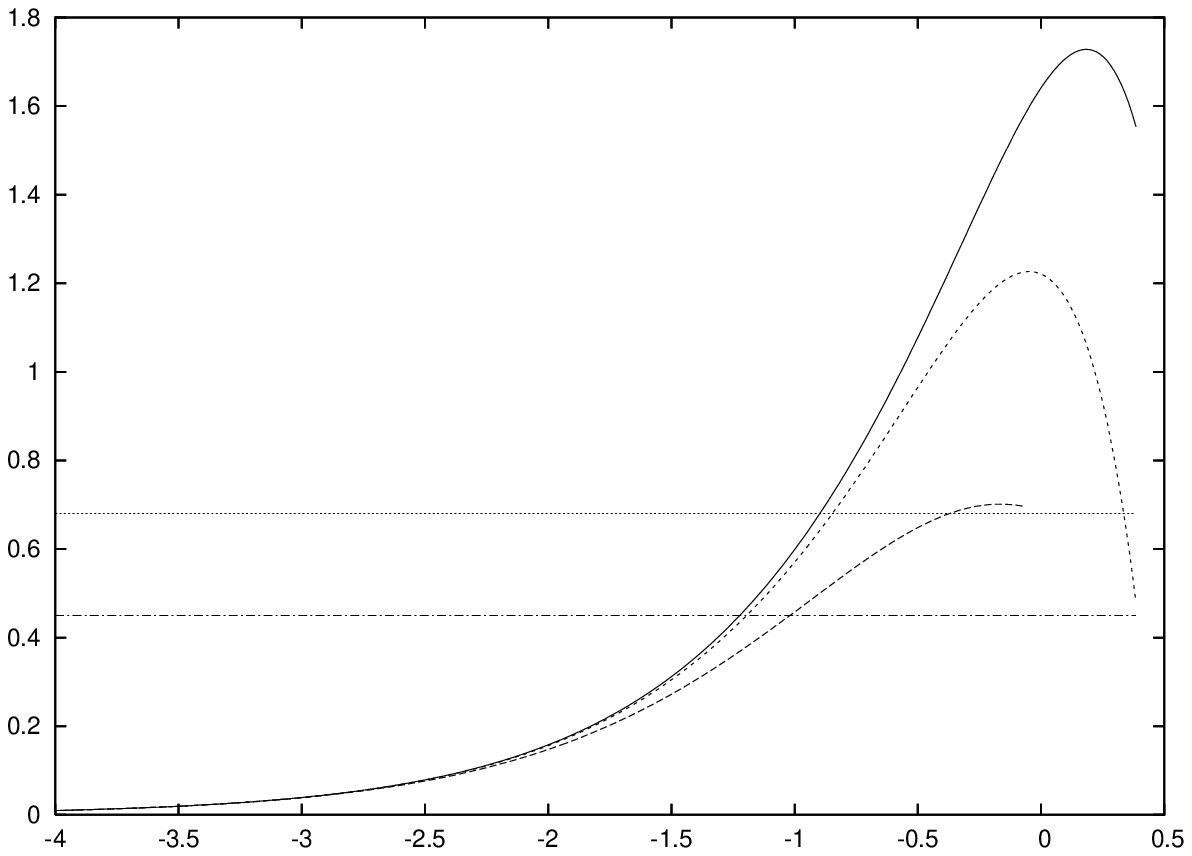}}b)\resizebox{2.8in}{1.8in}{\includegraphics{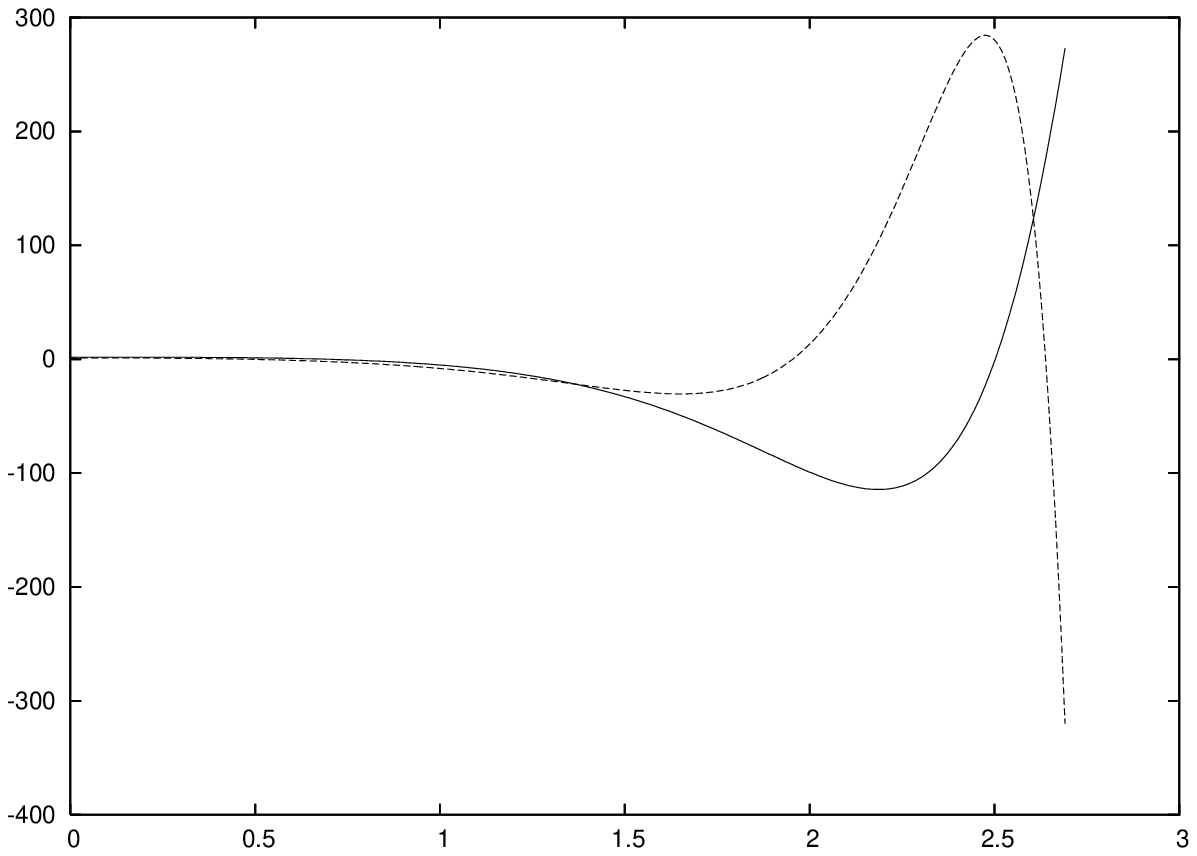}}\end{center}
Figure 5: a) The level $2$ approximation to the rolling of $\phi$ in the lightcone basis, graphed in comparison to the $\phi,B_+$ solution at level $\tilde{2}$, at the bottom, and the level $0$ tachyon solution the center of mass basis, at the top. b) The level $2$ approximation to $\phi$ (dotted line) for later times $0<t<2.7$, graphed in comparison to level $0$ (solid line). We have not graphed the $\phi,B_+$ solution out here since the series does not converge.\end{figure}
 
Thus it seems possible to approximate solutions in one basis by making an appropriate truncation in the other. However, this brings up a natural question: Precisely in what sense does level truncation in $\a^+_{-2n}$ modes in the lightcone basis converge to the equations of motion eq.\ref{eq:tach_EOM} for the tachyon in the center of mass basis? To explain this, consider the level $\tilde{2}$ equations of motion  for $\phi,B_+$, eq.\ref{eq:2p_EOM}. To compare this with eq.\ref{eq:tach_EOM}, we substitute in center of mass fields via the transformation law $\phi=\phi^\cm,B_+ = B_+^\cm - \frac{1}{\sqrt{2}}\partial_+\phi^\cm$ and suppose $B_+^\cm$ can be ignored. Then eq.\ref{eq:2p_EOM} results in the following approximate equations for the pure tachyon $\phi^\cm$:\begin{eqnarray}0=(\partial_+\partial_--1)\phi^\cm &+& g\kappa\left[(\phi^\cm)^2-\frac{1}{\sqrt{2}}(G_{\phi\phi B_+}-G_{B_+\phi\phi})\phi^\cm\partial_+\partial_-\phi^\cm\right.\\ &\ &\ \ \ \ \ \left.+\frac{1}{\sqrt{2}}G_{B_+\phi\phi}\partial_+\phi^\cm\partial_-\phi^\cm+...\right] \label{eq:EOM_approx}\end{eqnarray} where $+...$ are nonlinear terms with two $\partial_+$ derivatives, which we will not need for the sake of the present discussion. Since the nonlinear terms in eq.\ref{eq:EOM_approx} contain only up to $2$ lightcone derivatives in the interaction, whereas eq.\ref{eq:tach_EOM} contains an infinite number, it is clear that they must be related by some sort of polynomial expansion of the infinite differential operator $\exp(-V_{00}\partial_+\partial_-)$. Expanding the derivatives in the interaction term out to first order in $\partial_+$, eq.\ref{eq:tach_sol} becomes, \begin{eqnarray}0=(\partial_+\partial_--1)\phi^\cm + g\kappa\left[(\phi^\cm)^2 - 4V_{00}\phi^\cm\partial_+\partial_-\phi^\cm -2V_{00}\partial_+\phi^\cm\partial_-\phi^\cm+...\right]\nonumber\end{eqnarray} This leads to the identifications, $$ 4V_{00}\sim\frac{1}{\sqrt{2}}(G_{\phi\phi B_+}-G_{B_+\phi\phi})\ \ \ 2V_{00}\sim -\frac{1}{\sqrt{2}}G_{B_+\phi\phi}$$ Noting that $$ G_{\phi\phi B_+}= \frac{1}{\sqrt{2}}(2V_{02}^{33}-V_{02}^{13}-V_{02}^{23})\ \ \ \ \ G_{B_+ \phi\phi}=-G_{\phi\phi B_+}$$ these equations amount to the identification, $$ V_{00} \sim \frac{1}{2\sqrt{2}}(2V_{02}^{33}-V_{02}^{13}-V_{02}^{23})$$ Noting that $V^{33}_{0n}+V^{13}_{0n}+V^{23}_{0n}=0$, this can be rewritten, $$\ln\frac{27}{16}\sim-3\left(-\frac{1}{\sqrt{2}}V_{02}^{33}\right) $$ This is the first approximation to the exact identity\cite{Erler}, \begin{equation}\ln\frac{27}{16} = -3\sum_{n=1}^{\infty}\frac{(-1)^n}{\sqrt{2n}}V_{0,2n}^{33}\label{eq:identity}\end{equation} So it seems there are two levels of approximation going on. The first is polynomial approximation to the infinite order differential operator $\exp(-V_{00}\partial_+\partial_-)$; the second is an approximation to the coefficients of this polynomial through eq.\ref{eq:identity}. Using this knowledge we can get an estimate for how a solution including $\a^+_{-2n}$ modes up to level $\tilde{n}$ converges to the center of mass solution, eq.\ref{eq:tach_EOM}. Define $P_k(x)$ to be the $k$th order polynomial approximation to $e^x$, expanded around $x=0$. Roughly speaking, the level $\tilde{2k}$ approximation to eq.\ref{eq:tach_EOM} is, $$ 0 = (\partial_+\partial_--1)\phi^\cm + g\kappa P_k(-V_{00}\partial_+\partial_-)\left[P_k(- V_{00}\partial_+\partial_-)
\phi^\cm\right]^2 +...$$ Here $+...$ includes interactions with fields at higher levels and some higher derivative $(\phi^\cm)^2$ nonlinear terms, which we will ignore. Also, we have assumed that $V_{00}$ has been adequately approximated by the identity eq.\ref{eq:identity}. Making a perturbative ansatz, the solution is, $$ \phi_n^\cm = -\frac{g\kappa P_k(-\half
V_{00}n^2)}{n^2-1}\sum_{j=1}^{n-1}\phi_j^\cm P_k(-\half
V_{00} j^2) \phi_{n-j}^\cm P_k(-\half V_{00}(n-j)^2)$$ It is clear that for large enough $n$ the coefficients in this solution will look very different from eq.\ref{eq:tach_sol}. Only for the first few modes, where $P_k(-\half V_{00}n^2)$ is a good approximation to $e^{\frac{1}{2}V_{00}n^2}$ do the coefficients look similar. So, if we want a good approximation to $\phi_n$ up to some integer $n$, how many $\a^+_{-2n}$ modes do we need to include in the lightcone basis? Assuming we want to reproduce the $\phi_n^\cm$ to an accuracy of $10\%$, the answer is shown in Table 1. Roughly speaking, as we consider higher modes we must increase the level of truncation quadratically to get answers similar to those in the center of mass basis. Similar comments apply if we want to approximate solutions in the lightcone basis by level truncation in the center of mass basis. 
\begin{figure}[top]
\begin{center}
\begin{tabular}{|r|r|r|r|r|r|r|r|r|}\hline $n$, ($\phi^\cm_n$) & 2 & 3 & 4 & 5 & 6 & 7 & ... & $n$ \\ \hline level $2k$ & 4 & 8 & 14 & 22 & 32 & 44 & ... & $n(n-1)+2$ \\ \hline \end{tabular} \end{center} Table 1: This table shows what level in $\a^+_{-2k}$ excitations we need to include to get $\phi_n^\cm$s, up to some $n$, resembling those in the center of mass basis within $10\%$. \end{figure}

So we have studied two types of time dependent solutions describing the decay of an unstable D-brane: those arising naturally in the center of mass basis and those arising naturally from the lightcone basis. We have seen that, with a suitable truncation and definition of the equations of motion, it is possible to approximate the ``natural'' solution in one basis by fields in the other. However, this brings up an obvious question: which solution is more representative of the {\it actual} time dependent behavior of the string field as the D-brane decays? At this stage that we have seen no reason to favor one type of solution over the other. The difference between the lightcone and center of mass solutions really comes down to the fact that they solve {\it different} equations of motion, despite the fact that the field content of the solutions is identical. Indeed there is an intrinsic ambiguity in defining the equations of motion in the $\a^+_{-2n}$ truncation scheme. In order to derive equations of motion for $\a^+_{-2n}$ fields, we must vary the action with respect to $\a^-_{-2n}$ fields, but we do not separately account for the equations that the $\a^-_{-2n}$ fields must satisfy---rather, we have simply set $\a^-_{-2n}$ fields to zero. This means that the equations of motion in the $\a^+_{-2n}$ truncation scheme are really only defined up to terms proportional to the $\a^-_{-2n}$ equations of motion. This ambiguity accounts of the profound difference between the solution at level $0$ in the center of mass basis and that at level $\tilde{0}$ in the lightcone basis.

To resolve this ambiguity one might propose to include $\a^-_{-2n}$ fields and solve their equations of motion. We will do essentially this (also including other fields out to level $\tilde{2}$) in the next section. However, we do not believe this is likely to give a more realistic picture of D-brane decay. The reasons are twofold: First, $\a^-_{-2n}$ excitations create spurious solutions at the linear level in the lightcone basis---as we saw in section 3. Second, more fundamentally, $\a^-_{-2n}$ fields infect solutions with negative energies. We will have more to say on this in the next section, but it suffices to remark that these unphysical negative energies can only be projected out with a consistent imposition of constraints.  That being said, we believe that the $\phi^3$ solution at level $\tilde{0}$ probably more closely resembles the true decay process. The reason is that eq.\ref{eq:level0} possesses two features we know must be true of the exact theory: it is local in lightcone time and has a Hamiltonian free of all instabilities except the physical one associated with the tachyon.

\section{Solutions at level $(\tilde{2},\tilde{4})$ }
We now turn towards finding time dependent solutions representing
the decay of the D-$25$ brane vacuum, including all relevant
fields up to level $(\tilde{2},\tilde{4})$. For our purposes it
will be sufficient to restrict to solutions which depend only on
$\tilde{x}^+$ (time) and $x^-$, but not on the remaining $24$
transverse directions. To find the most efficient consistent
truncation at level $\tilde{2}$, we require that the string field
$|\Psi\rangle $ satisfy the following properties:
\bigskip

 1) $|\Psi\rangle$ is real and has ghost number 1.

 2) $b_0|\Psi\rangle=0$

 3) $|\Psi\rangle$ is twist even, i.e. only has excitations at even levels.

 4) $|\Psi\rangle$ is a singlet under the transverse subgroup
$SO(24)$ of the Lorentz group.
\bigskip

\noindent Condition 1 is simply a property of a physical string
field. Condition 2 is our choice of gauge. Condition 3 follows
from the well-known fact that twist odd fields only appear
quadratically in the string field theory action, so setting them
equal to zero solves their equations of motion identically.
Condition 4 follows from symmetry; if we require $|\Psi\rangle$ to
be a singlet initially, it will remain a singlet under time
evolution since the solutions we study have zero transverse
momentum. Imposing these conditions, we find that the string field
at level $\tilde{2}$ takes the form,
\begin{eqnarray}|\Psi\rangle &=& \int dk \left[\phi(k)+D_\parallel(k)\a^+_{-1}\a^-_{-1} +\frac{1}{\sqrt{48}}D_\perp(k)\a^i_{-1}
\a^i_{-1} + \frac{1}{\sqrt{2}}(D_{++}(k)\a^+_{-1}\a^+_{-1}+
D_{--}(k)\a^-_{-1}\a^-_{-1}) \right.\nonumber\\ &\ &\ \
+\left.\frac{i}{\sqrt{2}}(B_+(k)\a^+_{-2} + B_-(k)\a^-_{-2})+
\beta(k)c_{-1}b_{-1} \right]|k\rangle'\label{eq:l2_fields}\end{eqnarray} The eight spacetime fields above are expressed in the momentum
representation and are real in the momentum space sense, e.g.
$\overline{\phi(k)}=\phi(-k)$. Actually, there is another
simplification coming from the fact that we include interactions
only up to level $\tilde{4}$: we can consistently set $D_{++}=0$, and as a result $D_{--}$ decouples from the dynamics.  We can choose $D_{++}=0$ since the only nonlinear term in the $D_{++}$ equation of motion is proportional to $\phi D_{++}$; the other possible term $B_+B_+$ only enters at level $(\tilde{2},\tilde{6})$. By simple matching of Lorentz indices, it is easy to see that all other couplings are ruled out by the absence of $\partial_+$ derivatives in the vertex. Also by matching Lorentz indices it is possible to see that $D_{--}$ decouples when $D_{++}=0$, since the only way $D_{--}$ effects the dynamics is through a $D_{--}D_{++}$ term in the equation of motion for $\phi$; other couplings do not enter until $(\tilde{2},\tilde{6})$. Hence we can set $D_{++}=0$ and ignore $D_{--}$, bringing the total to six relevant spacetime fields at level $(\tilde{2},\tilde{4})$.

A natural approach to finding a solution is to expand the fields in modes $e^{\sqrt{2}nt}$, as we did in the last section. This method has the advantage of giving a spatially homogenous solution, like Sen's rolling tachyon\cite{Sen}.  However, we have found that the radius of convergence of this solution at level $(\tilde{2},\tilde{4})$ is not sufficiently large to follow the interesting dynamics---the series diverges before the tachyon turns around for the first time. One might attempt to perform some sort of Pad{\'e} resummation, but we have found this to be unreliable. 

However, in the lightcone basis we have an obvious alternative: we can make use of the fact that the theory possesses an initial value formulation. In particular, we can specify initial conditions for the fields at $\tilde{x}^+=0$, and then integrate the equations of motion to generate a full time dependent solution. To see how this is done, let us write the equations of motion for the eight fields eq.\ref{eq:l2_fields} in an abbreviated notation, \begin{eqnarray} 0&=&(\partial_+\partial_--1)\phi + \sqrt{2}\partial_+B_-+g\phi^{\mathrm{int}}[\phi,B_+,F_i]\nonumber\\
0&=&(\partial_+\partial_- +1)B_+ + \sqrt{2}\partial_+\phi + g B_+^{\mathrm{int}}[\phi,B_+,F_i]\nonumber\\
0&=&(\partial_+\partial_- +1)F_i +gF_i^{\mathrm{int}}[\phi,B_+,F_i]\label{eq:l2_EOM}
\end{eqnarray} where $F_i = B_-,D_\perp,D_\parallel,\beta,D_{--},D_{++}$, and $\phi^{\mathrm{int}}...$ denote the nonlinear terms in the equations of motion. There are over a hundred of these terms at level $(\tilde{2},\tilde{4})$---we will not write them explicitly---but we emphasize that they contain no lightcone time derivatives $\partial_+$. We can integrate the equations of motion eq.\ref{eq:l2_EOM} stepwise on a lattice, defining, $$\tilde{x}^+ = n\epsilon\ \ \ \ \phi_n(x^-)=\phi(n\epsilon,x^-)\ \ \ \ B_{+,n}(x^-)=B_+(n\epsilon,x^-)\ \ \ \ F_{i,n}(x^-)=F_i(n\epsilon,x^-)$$ where $\epsilon$ is a sufficiently small number. We can then use eq.\ref{eq:l2_EOM} to generate the fields at the $n+1$st time from the fields at the $n$th time:
\begin{eqnarray}F_{i,n+1}(x^-) &=& F_{i,n}(x^-)-\epsilon\int_{x^-_0}^{x^-}da\left[F_{i,n}(a)+gF^{\mathrm{int}}[\phi_n,B_{+,n},F_{i,n}](a)\right]\nonumber \\ \phi_{n+1}(x^-)&=&\phi_n(x^-)-\epsilon\int_{x^-_0}^{x^-}da\left[\sqrt{2}\frac{B_{-,n+1}(a)-B_{-,n}(a)}{\epsilon}+g\phi^{\mathrm{int}}[\phi_n,B_{+,n},F_{i,n}](a)-\phi_n(a)\right]\nonumber\\
B_{+,n+1}(x^-)&=&B_{+,n}(x^-)-\epsilon\int_{x^-_0}^{x^-}da\left[\sqrt{2}\frac{\phi_{n+1}(a)-\phi_n(a)}{\epsilon}+gB_+^{\mathrm{int}}[\phi_n,B_{+,n},F_{i,n}](a)+B_{+,n}(a)\right]\nonumber\\ \label{eq:disc}\end{eqnarray}
In these expressions we have already fixed a boundary condition which requires the fields to be {\it constant} in time $\tilde{x}^+$ along the light-like surface $x^-=x^-_0$. This choice of boundary condition corresponds to the familiar $p_-=0$ ambiguity occurring in any relativistic field theory whose initial value formulation is constructed in the lightcone frame. For most of our discussion we will take $x^-_0$ to be sufficiently far in the past that we can assume that the fields vanish. Note that these formulas must be solved in the order presented: The first formula gives the $F_i$ at $n+1$st time given the fields at $n$; the second gives $\phi$ at $n+1$ given the fields at $n$ and $B_-$ at $n+1$ (which we found in the first formula); the third gives $B_+$ at $n+1$ given the fields at $n$ and $\phi$ at $n+1$ (which we found from the second formula). Thus, taking $\epsilon$ to zero, we generate a solution by specifying the field configuration at $\tilde{x}^+=0$.

This technique gives us a way of generating a wide class of solutions from generic initial conditions. Unfortunately, except for very special cases, the resulting solutions are not homogenous, i.e. depend on both $\tilde{x}^+$ and $x^-$. Nevertheless, we have found that generally these types of solutions look similar to the homogenous (perturbative) solution when the radius of convergence of the latter allow them to be usefully compared---at least for initial conditions that are a sufficiently ``small'' perturbation of the open string vacuum.  

For definiteness, in most of our discussion we will choose the initial conditions, \begin{equation}\phi(x^-)=\frac{H}{4\cosh^2\half x^-}\ \ \ \ B_+(x^-) = -\frac{H}{2\sqrt{2}}(1+\tanh\half x^-)\ \ \ \ F_i(x^-)=0 \label{eq:in_cond} \end{equation} where $H$ is ``sufficiently small'' (for our purposes we choose $H=.1$). For most purposes, this choice of initial conditions seems irrelevant; any small perturbation of the D-brane generates a similar decay as long as the tachyon is pushed towards the closed string vacuum. Of course, if the initial conditions are too ``large,'' the resulting solution will depend more sensitively on the exact profile of the fields at $\tilde{x}^+=0$. For the most part, however, we are interested in generic features of the decay resulting from small perturbations of the D-25 brane, and not on detailed features of the time evolution of certain large field excitations. Therefore ``small'' initial conditions will serve us well.   

That being said, eq.\ref{eq:in_cond} has some noteworthy advantages as a choice of initial conditions. First, for early times $t=\frac{1}{\sqrt{2}}(\tilde{x}^++x^-)$ eq.\ref{eq:in_cond}  generates a homogenous solution in the linear theory, and as a result the first wave-crest in the nonlinear solution coincides roughly with $t=0$. If, for example, we had chosen gaussian initial conditions, the first wave crest would advance at the speed of light along the line $x^-=0$. Still, however, the solution would possess all of the principal qualitative features we will discuss. The second advantage of eq.\ref{eq:in_cond} is that they solve the truncated constraints for string field theory in Siegel gauge. As mentioned earlier, a time dependent solution of the Siegel gauge equations of motion eq.\ref{eq:EOM} does not {\it a priori} represent a full solution to the string field equations $Q_B\Psi + g\Psi*\Psi=0$. The other ``half'' of the string field equations, not accounted for by the Siegel gauge equations of motion, amount to a set of constraints which the string field must satisfy at all times (particularly $\tilde{x}^+=0$), like Gauss's law in electrodynamics. In ref.\cite{Erler} the constraints were found to take the form, \begin{eqnarray}0=b_0(c(\p2)P^+(\p2)-i\pi_b(\p2)x^+(\p2)')Q_B\Psi\label{eq:lc_constr}\end{eqnarray} where $c(\p2),P(\p2),\pi_b(\p2),x(\p2)'$ are the $c$ ghost, string momentum, $b$ ghost momentum, and worldsheet derivative of the string position evaluated at the string midpoint, respectively. In principle, the constraints eq.\ref{eq:lc_constr} play an extremely important role in the theory: they restrict dynamics to a submanifold in phase space where the Hamiltonian is positive (modulo the tachyon). This essentially amounts to a field-theoretic restatement of the no-ghost theorem of first quantized strings. It is not difficult to see that constraints are necessary to give physically acceptable solutions. Even at the linear level and ignoring the tachyon, we can see from the Siegel gauge action (at level $\tilde{2}$), \begin{eqnarray} S_\mathrm{free} &=& \int dx\left[\half\phi(\half\partial^2-1)\phi + \half D_\parallel(\half\partial^2+1)D_\parallel +\half D_\perp(\half\partial^2+1)D_\perp\right.\nonumber\\
&\ &\ \ \ \left.+ D_{++}(\half\partial^2+1)D_{--} -B_+(\half\partial^2+1)B_- - \half\beta(\half\partial^2+1)\beta +\sqrt{2}\phi\partial_+B_- \right]\nonumber\end{eqnarray}
that the Hamiltonian is unbounded from below, containing 
unphysical ``ghost'' degrees of freedom like $\beta$ and $\frac{1}{\sqrt{2}}(B_+ +B_-)$ whose kinetic and potential energies are negative definite. Some extra conditions are necessary to ensure that the dynamics generated by the Siegel gauge equations of motion is not completely unstable. Thus it seems important that the initial conditions eq.\ref{eq:in_cond} satisfy the constraints and therefore set the solution on the right track. Unfortunately, however, we will find that the qualitative behavior of our solutions, in particular the unphysical negative energy instabilities we will encounter, are largely insensitive to the fact that eq.\ref{eq:in_cond} satisfies the constraints. The reason is that, at finite level, the constraints are inconsistent with the equations of motion; even if the string field satisfies the constraints initially, it will in general fail to upon time evolution at finite level. Indeed, the consistency of the constraints eq.\ref{eq:lc_constr} with time evolution in the exact theory is a subtle issue relying on certain nontrivial identities between the Neumann coefficients\cite{Erler} which only emerge properly in the infinite level limit. Later, we will have to deal with unphysical instabilities in a more direct fashion, ``freezing out'' problematic negative energy fields by hand.

So let us present the level $(\tilde{2},\tilde{4})$ time dependent solution derived from these initial conditions. Figure 6a shows the dynamics of $\phi$ as a function of $\tilde{x}^+,x^-$, and figure 6b shows more specifically the profile at $\tilde{x}^+=2.5$. At early times the solution does the expected thing; it rolls off the open string vacuum, bounces off the $\phi^3$ potential  wall and turns around. What happens after this is more disturbing: the tachyon rolls back over the open string vacuum into the unbounded side of the $\phi^3$ potential, but somehow feels the urge to turn around again and shoots quickly back towards the open string vacuum. At later times the solution becomes increasingly unstable, oscillating back and forth with increasing frequency and amplitude as shown in figure 6c. After a certain point the gradients become too large to follow the solution reliably, but we have no reason to believe that the dynamics ever stabilizes.

\begin{figure}[top]\begin{center}a)\resizebox{2.8in}{1.8in}{\includegraphics{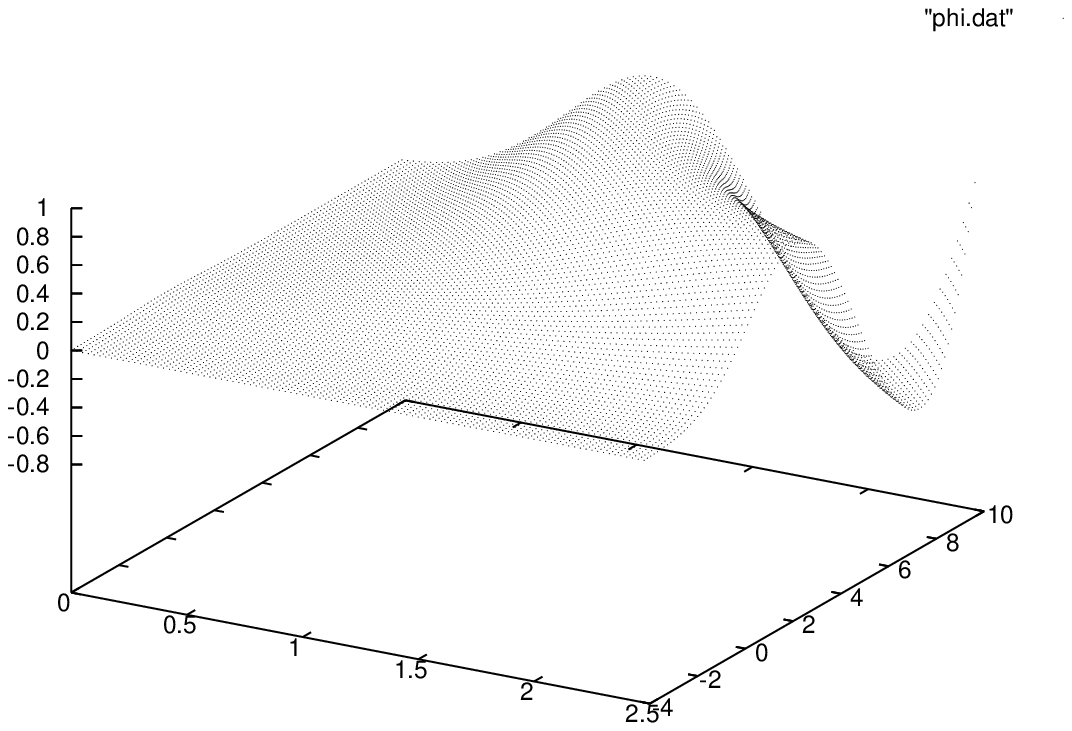}}b)\resizebox{2.8in}{1.8in}{\includegraphics{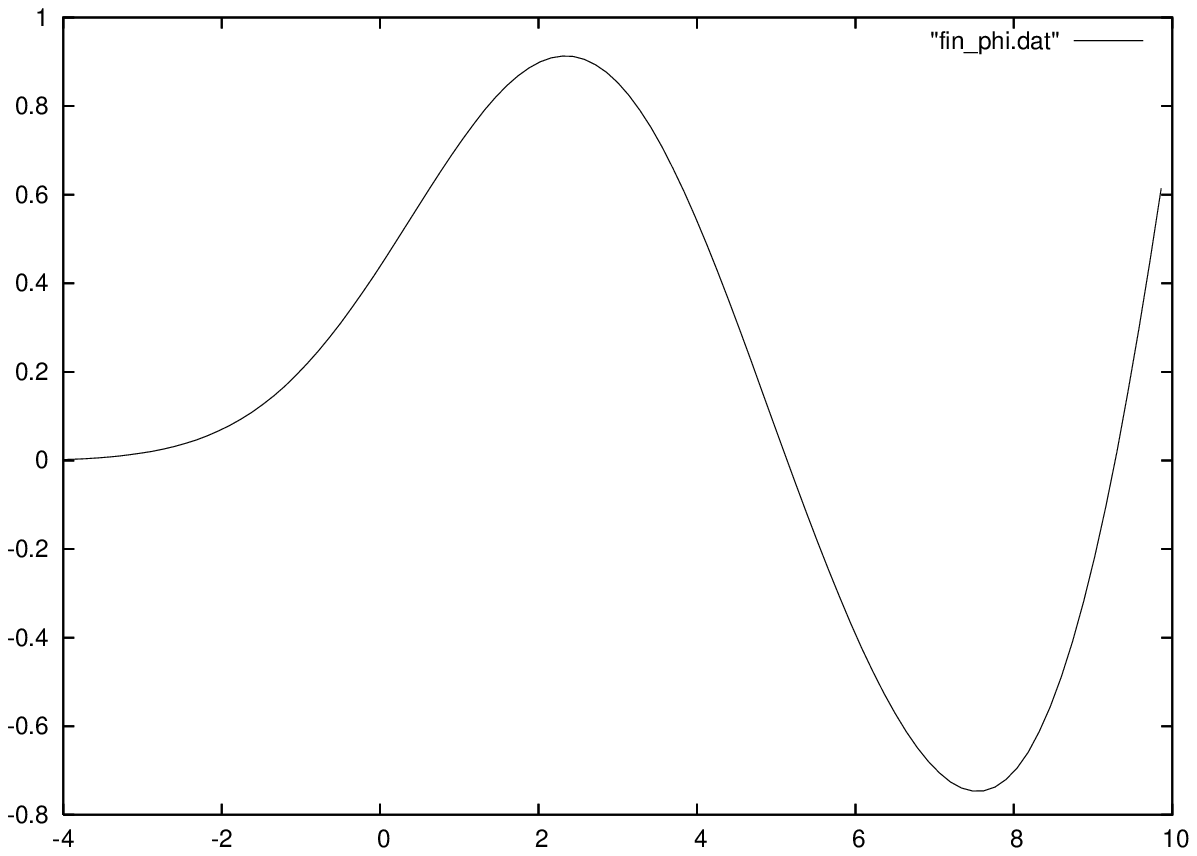}}
c)\resizebox{2.8in}{1.8in}{\includegraphics{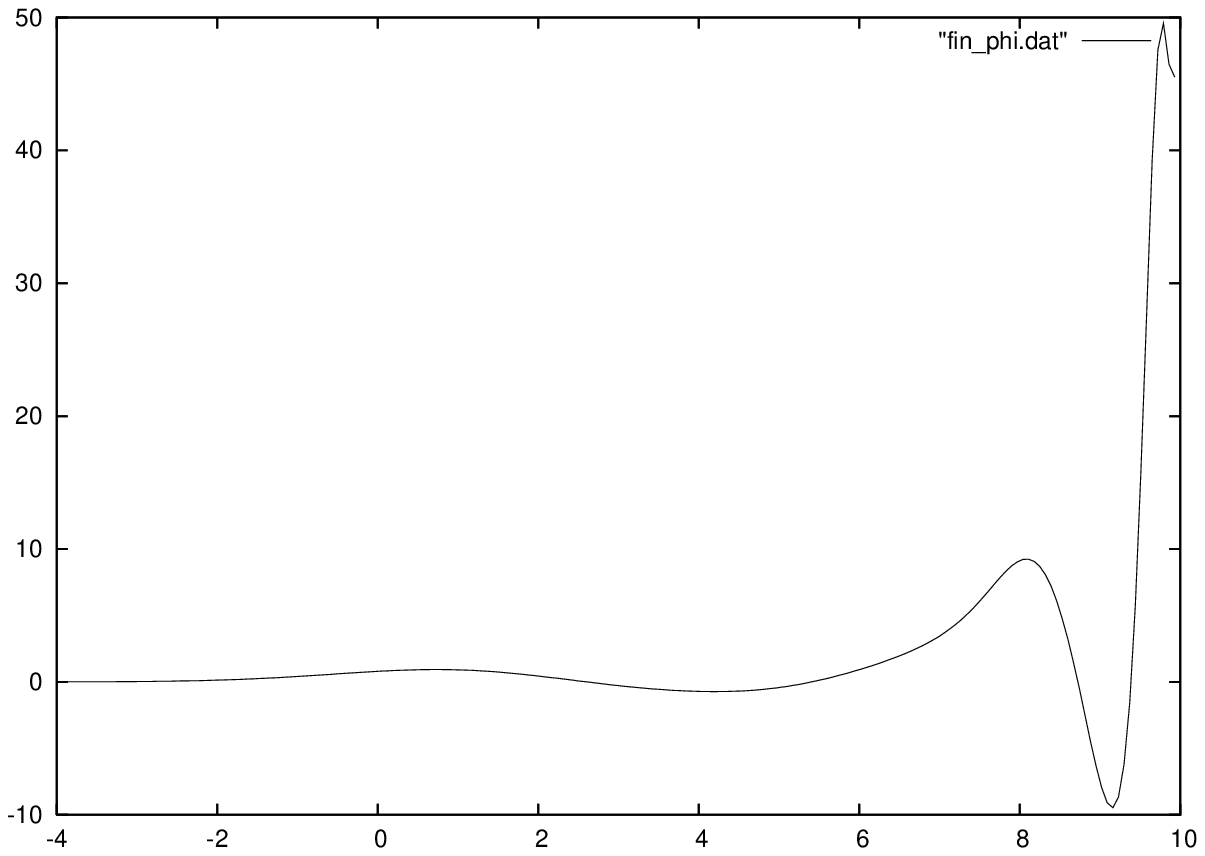}}\end{center}
Figure 6: a) The level $(\tilde{2},\tilde{4})$ approximation to the rolling of $\phi$ in the lightcone basis, graphed as a function of $\tilde{x}^+,x^-$ for $-4<x^-<10$ and $0<\tilde{x}^+<2.5$. b) The profile of $\phi$ at $\tilde{x}^+=2.5$, c) at $\tilde{x}^+=3.6$. \end{figure}

We hope, of course, that this solution does not in any way approximate the dynamics of D-brane decay. This erratic behavior clearly indicates that we are dealing with a system of equations which couple to unbounded negative energies. What mechanism is responsible for this erratic behavior? The problem seems closely related to the unphysical linear instability we discovered in section 3, where the linear motion of $B_-$ drives $B_+$ at its resonant frequency. In figure 7 we have graphed the profiles of $B_+,\phi,B_-$ at $\tilde{x}^+=2.5$; at its peak $B_+$ acquires a value approximately ten times that of other fields, strongly suggesting that $B_+$ is the principal victim of the instability. The fact that the linear motion of $B_-$ is the cause of the instability is supported by two bits of evidence. First, if the linear coupling of $B_-$ to the $\phi$ equations of motion is turned off, $B_+$ is not driven to large values and the resulting solution is much more well-behaved. Second, if the nonlinear couplings in the $B_-$ equations of motion are turned off, but we set nonvanishing initial conditions for $B_-$, the resulting purely linear motion of $B_-$ still causes the instability. However, nonlinear effects do seem to play an important role. For example, we saw in section 3 that including an extra field at level $\tilde{4}$, $\a_{-2}^+\a_{-2}^-|k\rangle'$, makes the linear instability of $B_+$ disappear. However, adding $\a_{-2}^+\a_{-2}^-|k\rangle'$ onto our solution does not make it more well-behaved. The reason, we think, is that by the time the instability sets in, the tachyon is far off shell and the cancellation of the unphysical $B_+$ instability no longer occurs. Another important nonlinear effect comes from the $\phi\partial_-B_+$ coupling in the $\phi$ equation of motion. If this term is set to zero, the instability does not occur in its present form.

\begin{figure}[top]\begin{center}\resizebox{2.8in}{1.8in}{\includegraphics{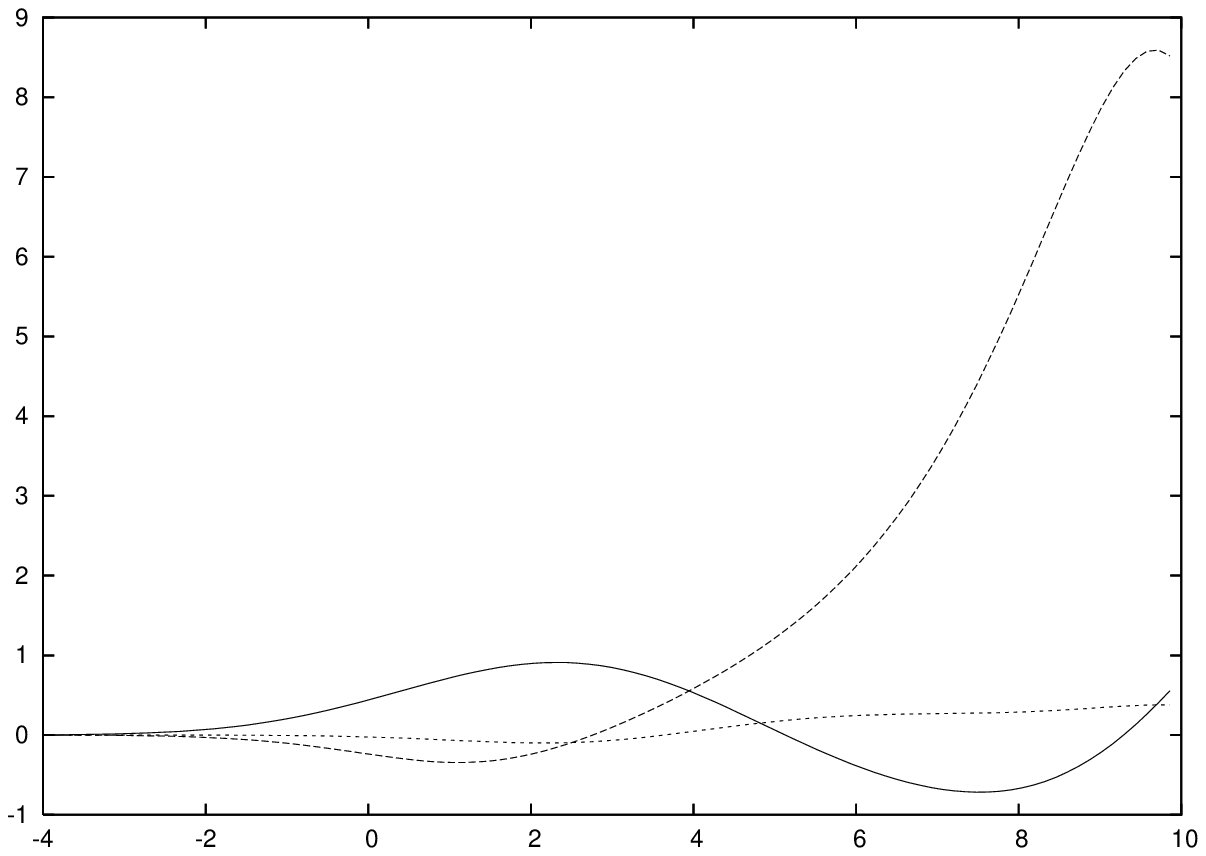}}\end{center}
Figure 7: The level $(\tilde{2},\tilde{4})$ profiles for $B_+$ (dashed line), $\phi$ (solid line), and $B_-$ (dotted line) at $\tilde{x}^+=2.5$.\end{figure}

Since the linear instability introduced by $B_-$ is an artifact of the level truncation scheme, it seems clear that the instability of the level $\tilde{2}$ solution is also an unphysical artifact of level truncation. However, limited at present to level $\tilde{2}$, it seems important to explore possible modifications of the equations of motion at this level which have a chance of painting a more realistic picture of D-brane decay. We propose to set all nonlinear couplings in the $B_-$ equation of motion to zero. Then, given our initial conditions eq.\ref{eq:in_cond}, this has the effect of setting $B_-=0$. Remarkably, with this modification alone the solutions we find will paint a consistent and believable picture of D-brane decay beyond level $\tilde{0}$.

With $B_-=0$, we have plotted in figure 8 the profile of $\phi$ at $\tilde{x}^+=2$ generated from the initial conditions eq.\ref{eq:in_cond}. For the purposes of comparison, we have graphed it alongside another profile generated from the equations of motion obtained upon setting $B_+=0$. When setting $B_+=0$, the equations of motion simplify to that of an ordinary cubic theory coupling three scalars $\phi,\beta, A=-\half D_\parallel = \frac{1}{\sqrt{48}}D_\perp$ with no derivative dependence in the interaction. The linear and nonlinear couplings of $B_+$ can be thought of as more characteristic of string theory, so the $B_+=0$ profile provides a useful gauge for identifying uniquely stringy behavior. We see two characteristic features emerge from figure 8: First, initially $\phi$ rolls somewhat more quickly towards the closed string vacuum relative to the $B_+=0$ profile. Second, $\phi$ does not become as large before turning around. Relative to the $B_+=0$ profile, the first peak is $6\%$ closer to the closed string vacuum. Both of these features are fairly insensitive to the choice of initial conditions.

\begin{figure}[top]\begin{center}\resizebox{3.8in}{2.8in}{\includegraphics{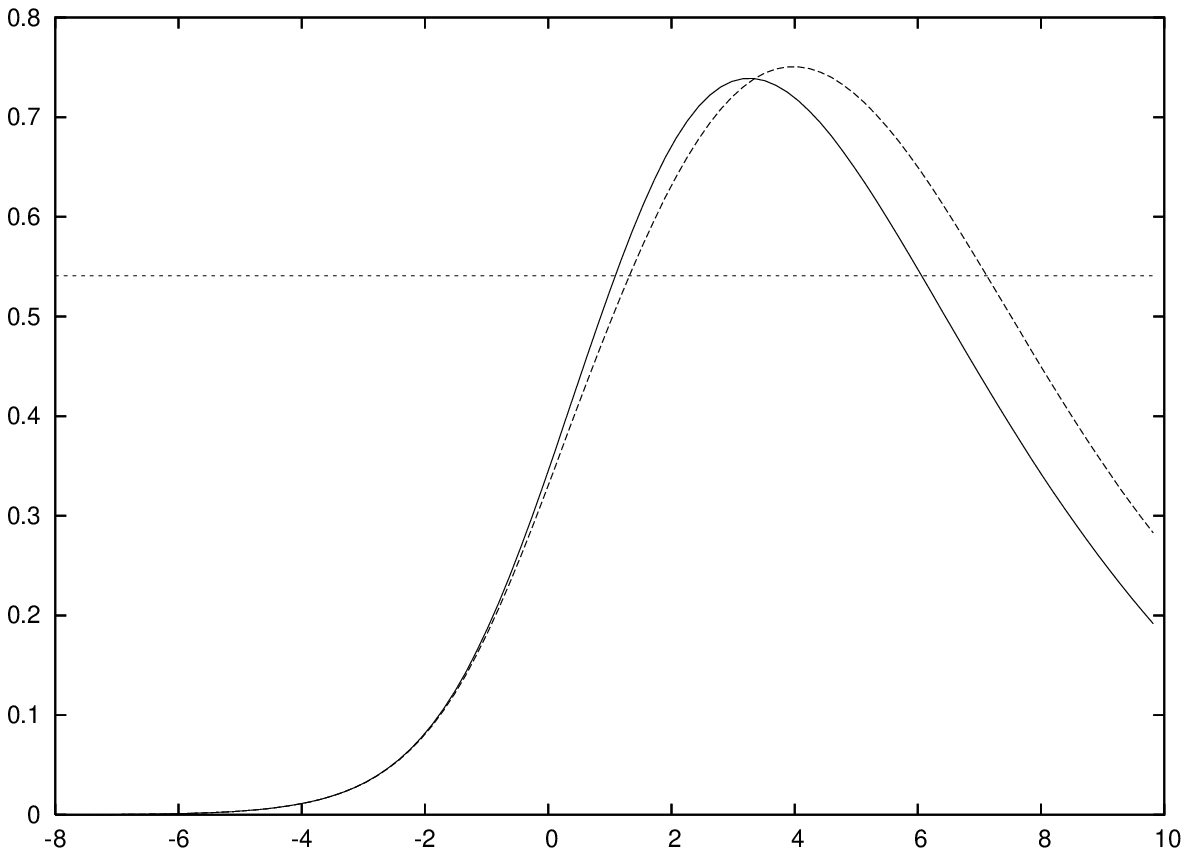}}\end{center}
Figure 8: Profile for $\phi$ (solid line) at $\tilde{x}^+=2$ including all fields except $B_-$ starting from the initial conditions eq.\ref{eq:in_cond}. The dashed line shows the analogous profile generated upon setting $B_+=0$. The horizontal lines denote the vev of $\phi$ at the closed string vacuum.\end{figure}

However, at this point it is worth mentioning that even having set $B_-=0$ there is another point of possible concern with the equations of motion: The field  $\beta$ has negative kinetic energy and may introduce further unphysical instabilities into solutions. It is worth seeing the nature of these instabilities, though we will see that they do not significantly effect the qualitative features we identified in figure 8. Suppose that we set initial conditions so that the string field is very close to the closed string vacuum, 
\begin{equation}\phi^\mathrm{vac}=.542,\ \ D_\parallel^\mathrm{vac} = -.0519\ \ \ D_\perp^\mathrm{vac} = .180\ \ \ \beta^\mathrm{vac}=-.173 \label{eq:in_cond2}\end{equation} but we displace $\phi$ away from the vacuum by a small amount $\Delta\phi=.01$. The resulting time dependent solution is plotted in figure 9. The tachyon oscillates back and forth around the stable vacuum, as expected, but the oscillations grow with time. This behavior is directly caused by the dynamics of $\beta$; fixing $\beta$ to its value at the closed string vacuum, the oscillations no longer grow with time. 

\begin{figure}[top]\begin{center}a)\resizebox{2.8in}{1.8in}{\includegraphics{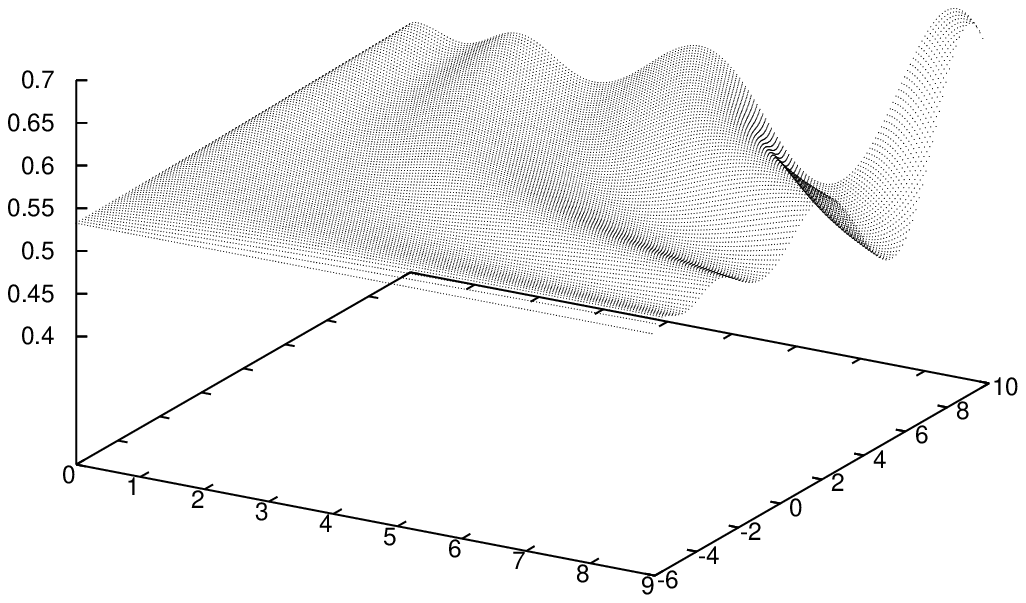}}b)\resizebox{2.8in}{1.8in}{\includegraphics{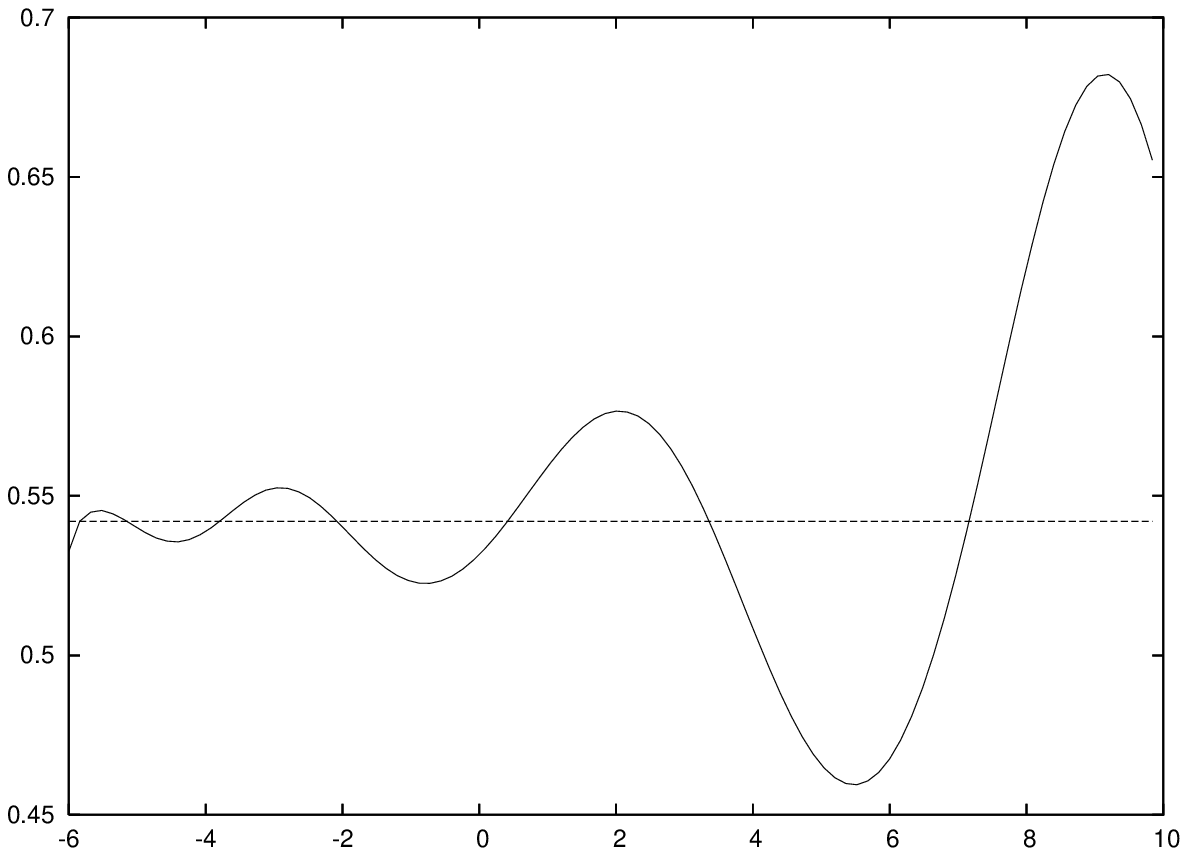}}\end{center}
Figure 9: Dynamics of $\phi$ at level $(\tilde{2},\tilde{4})$ (excluding $B_-$)
if we place the string field close to the closed string vacuum, plotted a) as a function of $0<\tilde{x}^+<9$, $-4<x^-<10$ and b) at $\tilde{x}^+=9$. The horizontal line denotes the vev of $\phi$ at the closed string vacuum.\end{figure}

We can try two things in an effort to deal with the negative energy instabilities of $\beta$. First, we can simply fix $\beta=0$, which is valid at least in the linear approximation. Truncating away $\beta$ however changes the location of the closed string vacuum,  $$\phi=.674\ \ \ D_\perp = .174\ \ \ \ \ D_\parallel = -.0503$$
and the vacuum energy density at this point is $-1.37$ times that of the D-25 brane tension. Starting from the initial conditions eq.\ref{eq:in_cond} we have generated a solution  with $\beta=B_-=0$ as shown in figure 10a. Figure 10b shows the profile of $\phi$ at $\tilde{x}^+=6$, graphed for comparison to the corresponding profile with $B_+=0$. Again we see that $\phi$ initially rolls faster to the closed string vacuum, but turns around $17\%$ closer to the closed string vacuum than it does with $B_+=0$. The second approach to dealing with $\beta$ is to set $\beta=-.173$, its value at the closed string vacuum, and look at small time dependent fluctuations around the vacuum. Using the initial conditions eq.\ref{eq:in_cond2}, at $\tilde{x}^+=4$ we get the $\phi$ profile shown in figure 11, alongside the analogous profile with $B_+=0$. Again $\phi$ falls faster to the closed string vacuum, but turns around early. The fact that these qualitative features emerge from three inequivalent truncations strongly suggests that this behavior not an artifact of our approximation scheme, and is largely independent of the negative energies of $\beta$. 

\begin{figure}[top]\begin{center}a)\resizebox{2.8in}{1.8in}{\includegraphics{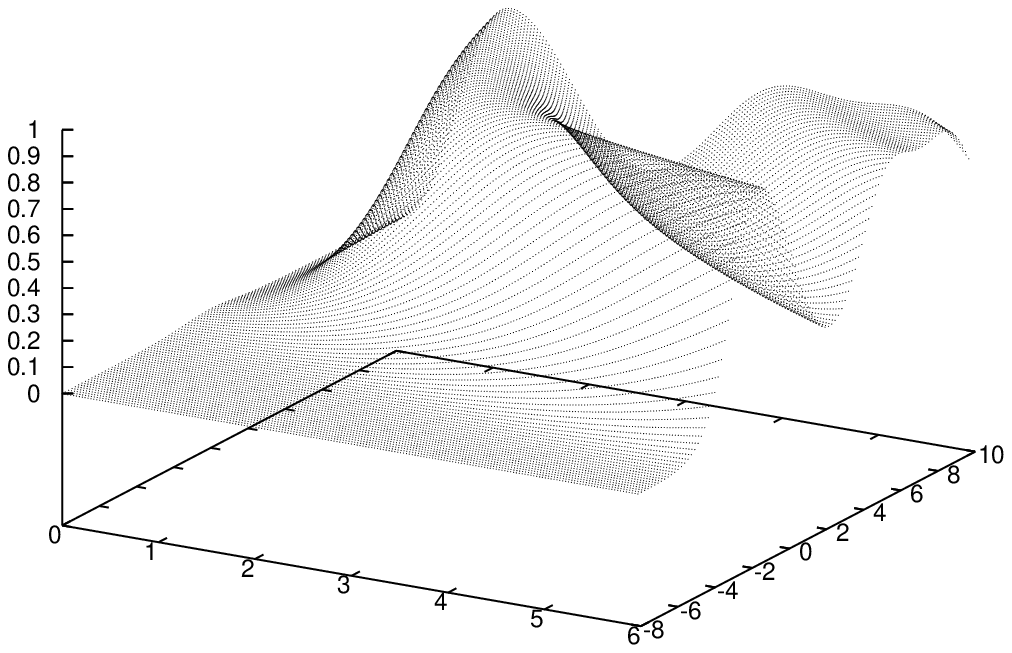}}b)\resizebox{2.8in}{1.8in}{\includegraphics{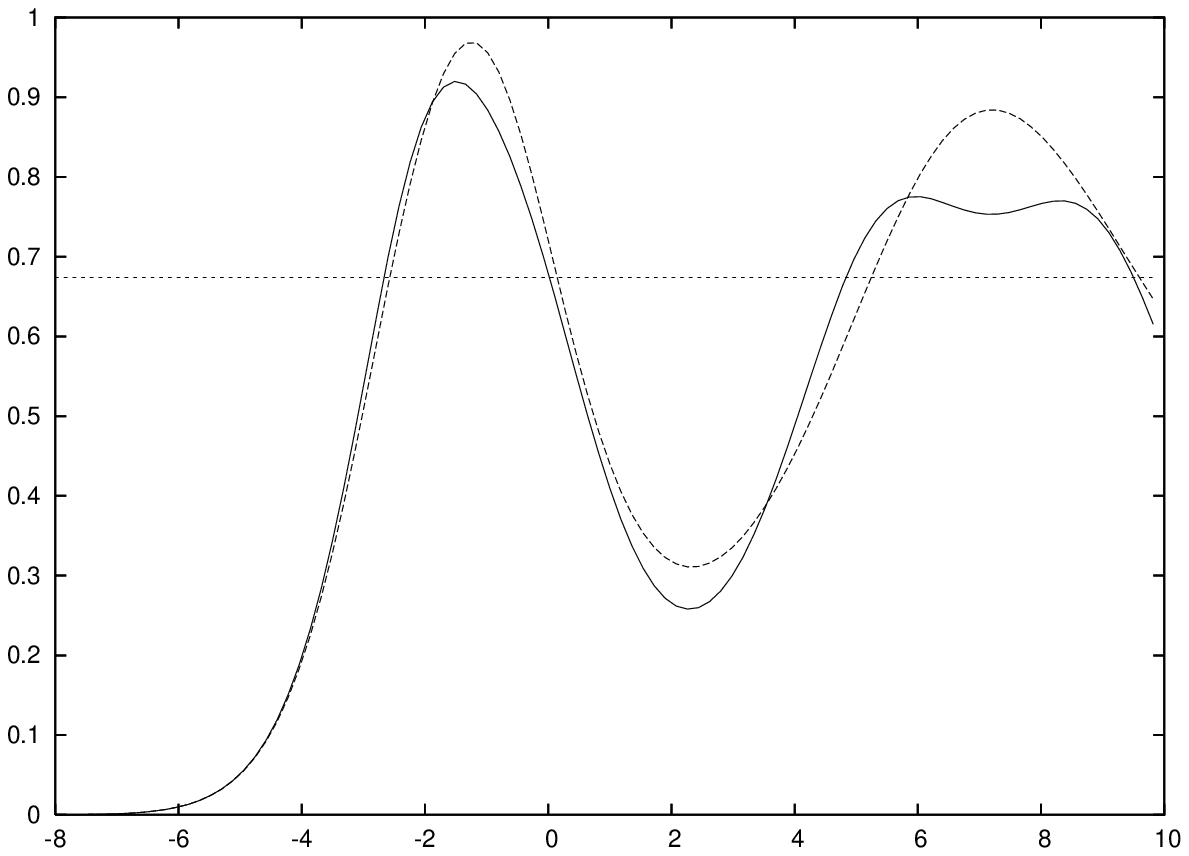}}\end{center}
Figure 10: Dynamics of $\phi$ at level $(\tilde{2},\tilde{4})$ excluding $B_-$ and $\beta$ a) as a function of $0<\tilde{x}^+6$, $-8<x^-<10$ and b) at $\tilde{x}^+=6$ (shown as solid line). The dashed line shows the same profile generated from the equations after setting $B_+=0$. The horizontal line denotes the vev of $\phi$ at the closed string vacuum.\end{figure}

\begin{figure}[top]\begin{center}\resizebox{2.8in}{1.8in}{\includegraphics{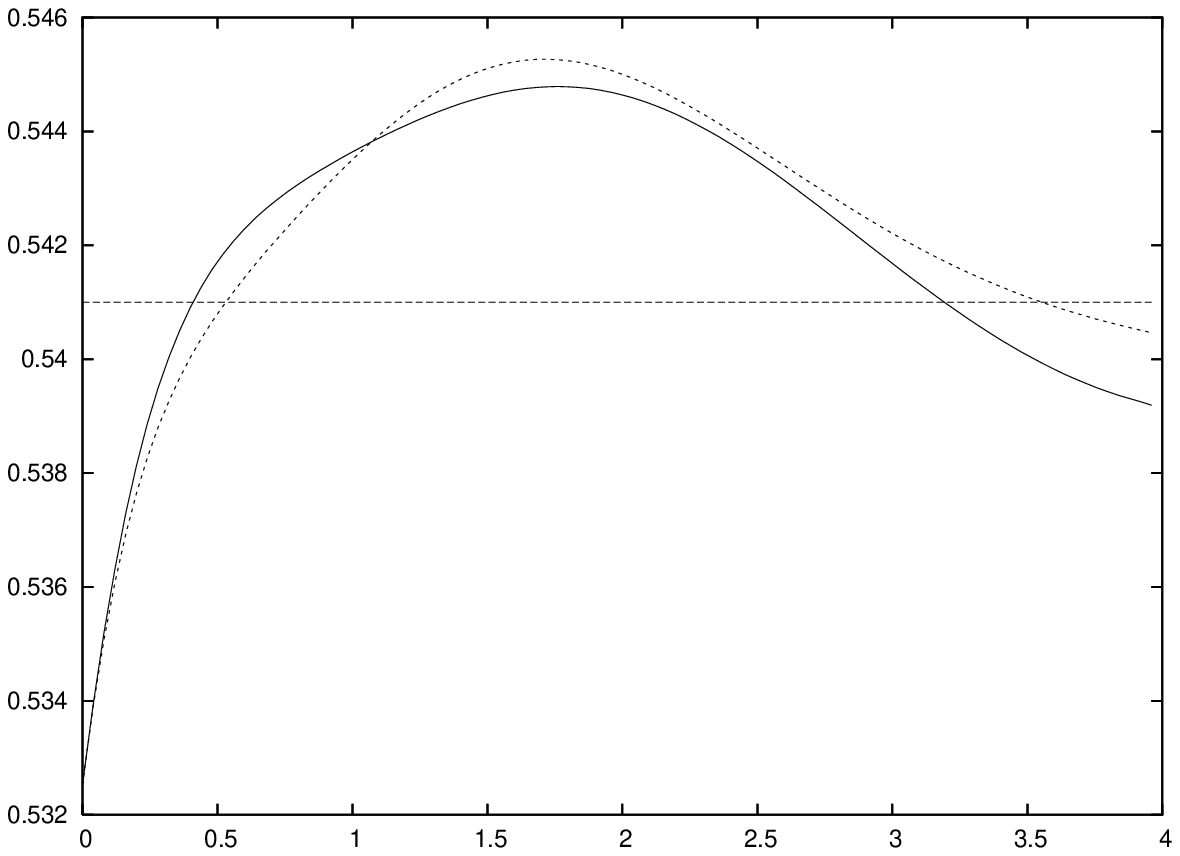}}\end{center}
Figure 11: Fixing $\beta=-.173, B_-=0$, the solid line shows small time dependent fluctuations around the closed string vacuum graphed at $\tilde{x}^+=4$ starting from the initial conditions eq.\ref{eq:in_cond2}. The dashed line shows the analogous profile obtained upon setting $B_+=0$.\end{figure}

Since $\phi$ turns around early in all of these solutions, it is tempting to hope that in the infinite level limit $\phi$ may not cross the closed string vacuum at all. Thus, the picture of D-brane decay in open string field theory would resemble Sen's ``rolling tachyon'' boundary conformal field theory solution, where the tachyon falls homogeneously off the unstable maximum towards the closed string vacuum, but does not cross over in finite time. Indeed, there are preliminary indications that at higher levels the height of the first peak of $\phi$ is further suppressed. For example, in figure 12 we have graphed the profile of $\phi$ at $\tilde{x}^+=6$ (starting from eq.\ref{eq:in_cond}) generated if we include the extra level $\tilde{4}$ field, $$\int dk \frac{i}{2}C_+(k)\a^+_{-4}|k\rangle$$
and set $\beta=B_-=0$. Now $\phi$ turns around even earlier, $7\%$ closer to the closed string vacuum than of we had set $C_+=0$ and  $23\%$ closer than if we had set $B_+=C_+=0$. However, before we jump to conclusions, it is worth mentioning that, from the perspective of open string field theory, it is not clear what it {\it means} for the string field to monotonically approach the closed string vacuum but not ``cross over'' in finite time. In string field theory the closed string vacuum corresponds to a single point in an infinite dimensional space of fields at zero momentum, and in the process of decay the string field might oscillate many times around the closed string vacuum but never actually cross over this point. To illustrate this, we have generated two parametric plots following the path of $\phi,D_\perp$, as a function of $x^-$ at $\tilde{x}^+=6$, as the D-brane decays. Setting the negative energy fields $B_-,\beta$ to zero, figure 13a shows the path followed if we truncate away $B_+$ and figure 13b shows the path including $B_+$. In 13a the string field oscillates regularly around the closed string vacuum; in 13b, $\phi$ does not become as large (as we saw before) but the oscillations around the vacuum seem much more chaotic. Indeed, it is not clear from these pictures how our solutions might be approaching Sen's ``monotonic'' rolling tachyon solution, or even how we would know it if they were. Still, the fact that $D_\perp$ oscillates more wildly for our solutions could have been expected, since the kinetic energy which would otherwise allow $\phi$ to roll further up the potential wall must be absorbed by more massive fields, causing them to oscillate more strongly. If the rolling tachyon solution exists in open string field theory, this could be the mechanism allowing the field to approach the closed string vacuum asymptotically---the energy originally stored in the tachyon and lightest mass fields gets absorbed by fields of arbitrarily large mass. 

\begin{figure}[top]\begin{center}\resizebox{2.8in}{1.8in}{\includegraphics{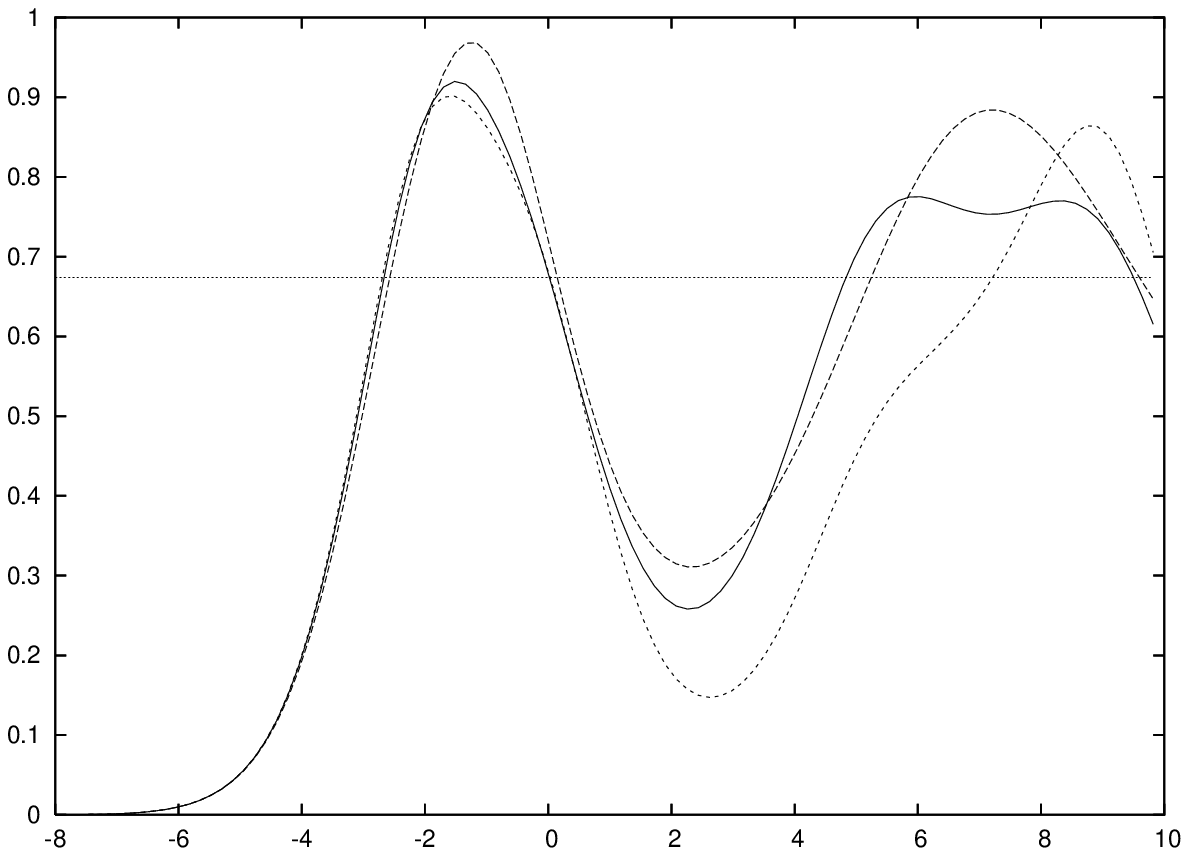}}\end{center}
Figure 12: Three profiles of $\phi$ at $\tilde{x}^+=6$ generated from the initial conditions eq.\ref{eq:in_cond}. The solid line includes all fields at level $\tilde{2}$ except $\beta,B_-$. The dotted line includes all fields at level $\tilde{2}$ except $\beta,B_-$, but also includes the level $\tilde{4}$ field $C_+$. The dashed line is the profile one obtains setting $B_+=C_+=0$. \end{figure}

\begin{figure}[top]\begin{center}a)\resizebox{2.8in}{1.8in}{\includegraphics{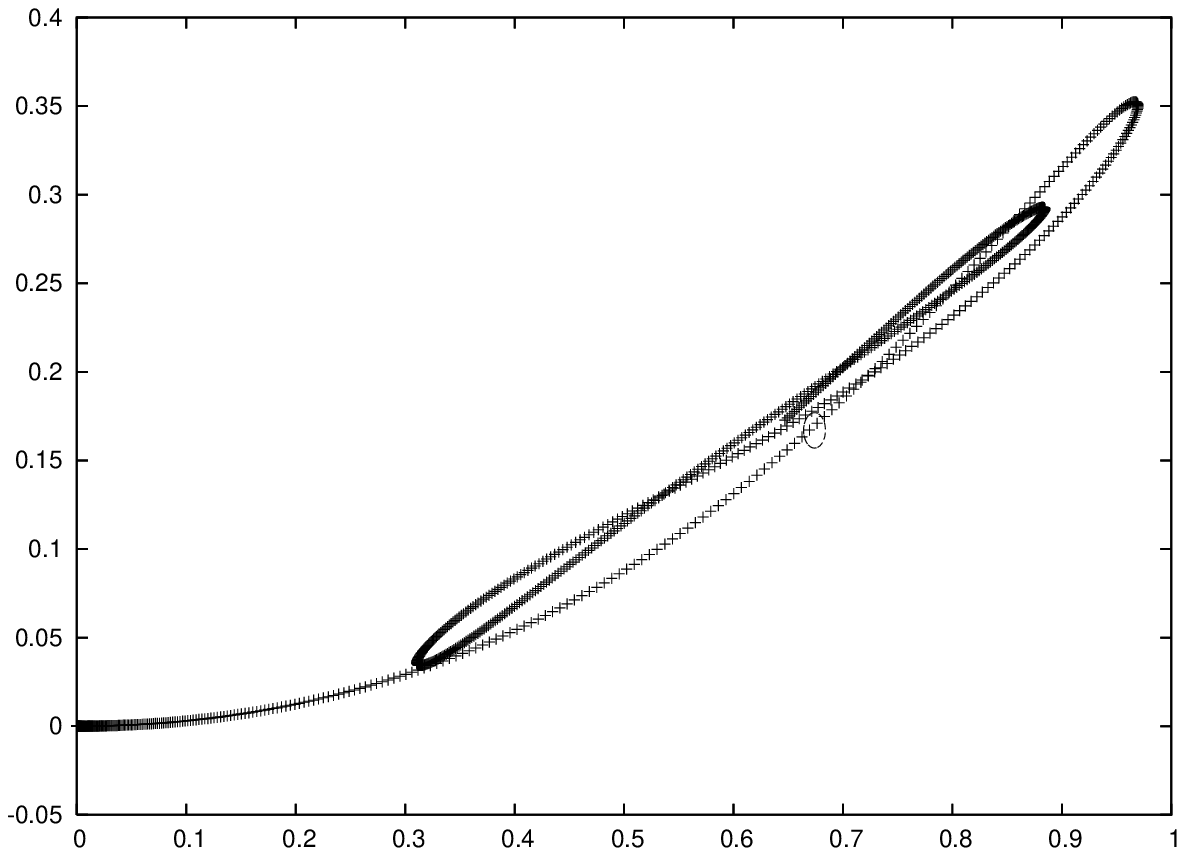}}b)\resizebox{2.8in}{1.8in}{\includegraphics{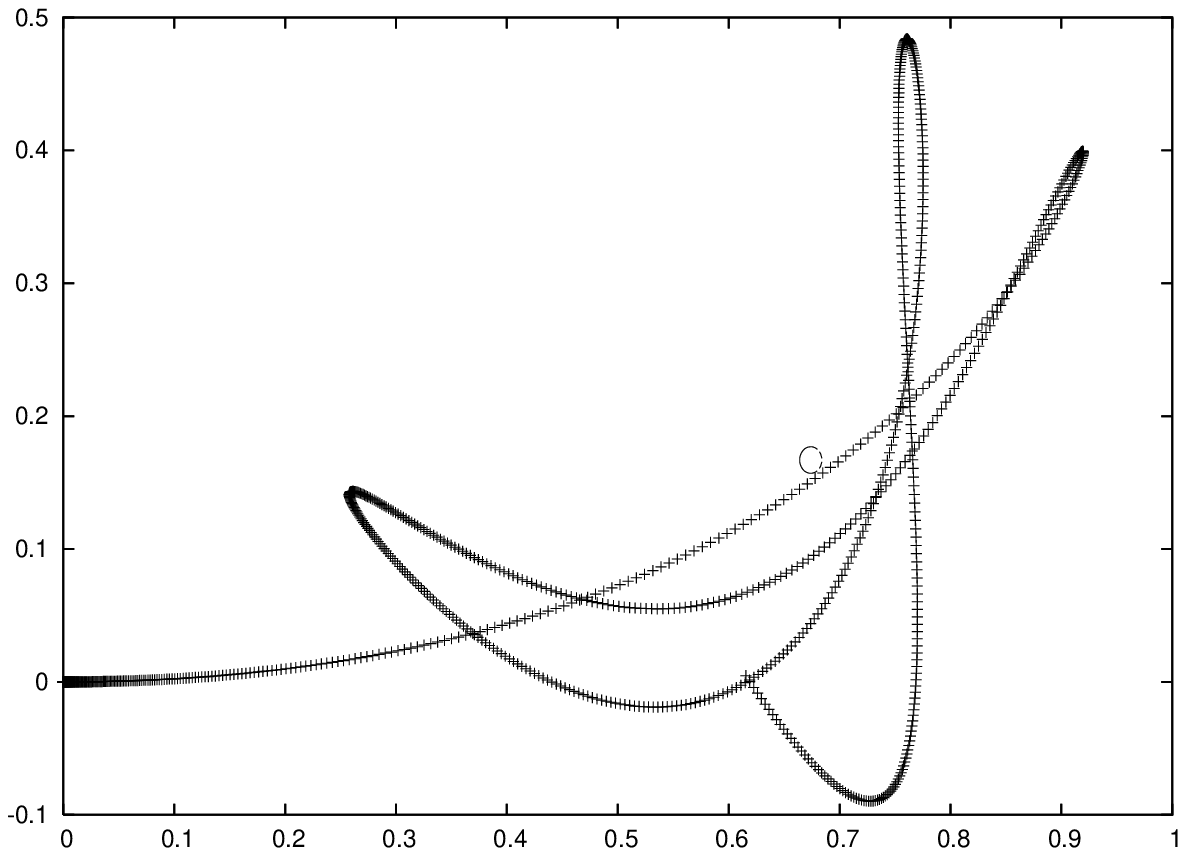}}\end{center}
Figure 13: Plots showing the dynamics of $\phi$ (horizontal axis) and $D_\perp$ (vertical axis) as parametric functions of $x^-$ at $x^+=6$. In both plots we truncated away the negative energy fields $B_-,\beta$. 12a shows the dynamics with $B_+=0$ and 12b shows the dynamics with $B_+\neq0$. The little circle in these figures shows the position of the closed string vacuum.\end{figure}

\section{Conclusion}

In this paper we have studied level truncation and time dependent solutions representing the decay of an unstable D25-brane in open string field theory. We introduced a basis of spacetime fields, the ``lightcone basis,'' for which the equations of motion are local in lightcone time and therefore admit a well-defined initial value formulation and a Hamiltonian free of higher derivative instabilities. We studied the effect of level truncation in the linear theory and investigated a modified level expansion which allowed us to understand the relationship between time dependent solutions in the lightcone basis and the center of mass basis, where the interaction vertex possesses an infinite number of time derivatives. Finally, we investigated time dependent solutions including all spacetime fields up to level $\tilde{2}$. We found that, due to spurious instabilities introduced by truncating the theory at the linear level, the solutions at level $\tilde{2}$ are unstable. However, by looking at a slightly modified set of equations of motion, we found much more regular solutions. The picture that emerges seems to indicate that, in comparison to solutions generated by only including the scalars at level $\tilde{2}$, the tachyon rolls faster towards the closed string vacuum initially but then rapidly decelerates and does not become as large before turning around. Though these results can only be regarded as a first step, they seem to indicate that it is at least possible that Sen's rolling tachyon solution can be realized in open string field theory.

Of course, it would seem highly desirable to take our analysis further, perhaps by including higher levels, to see of a more refined picture of the D-brane decay emerges. Unfortunately, at higher levels the equations of motion become exponentially more complicated and unphysical instabilities proliferate. In principle, if the constraints are consistently imposed and we go to high enough levels, the unphysical instabilities should work themselves out, but we have no understanding of how far we need to go for this to occur. Probably it is unreasonable to hope that a brute force approach to higher levels will yield a clearer picture of the decay. What is needed is a clearer analytic understanding of the constraints eq.\ref{eq:lc_constr}, particularly how they exorcise unphysical modes in the linear and nonlinear theory. The consistency of the constraints with time evolution is subtle, relying on nontrivial properties of the vertex in the infinite level limit. Indeed, it is not even clear that the closed string vacuum satisfies the constraints---though, barring some sort of anomaly, naively it does. Armed with a better understanding of constraints, it may be easier to identify the relevant approximations and take our analysis further. 

Of course, there are other techniques which may be worth exploring. One might be to follow the approach of ref.\cite{Zwiebach-Time}, inverse Wick rotating a marginal solution interpolating between a D1 and D0-brane compactified on a unit circle along $\tilde{x}^1=\frac{1}{\sqrt{2}}(\tilde{x}^+-x^-)$. Indeed, constructing time independent marginal and lump solutions carrying nonzero momentum along $\tilde{x}^1$ in the lightcone basis is of interest in its own right. Another possibly interesting approach is to find time dependent solutions using the technique of ref.\cite{Okawa} based on the regulated butterfly state, though so far this has only been applied towards approximating the closed string vacuum solution at zero momentum. One might also try to investigate time dependent solutions in the context of vacuum string field theory\cite{Bonora, Hata2}, where direct analytic understanding is possible. Unfortunately, it seems doubtful that the lightcone basis could be usefully applied in this context, since naively the action would contain no $\partial_+$ derivatives at all, rendering the dynamics trivial. Still, it may be interesting to reconcile the lightcone basis with some of the time dependent solutions constructed so far in vacuum string field theory.

I would like to thank D.Gross and W.Taylor for some useful discussions. This work was generously supported by the National Science foundation under Grant No.PHY99-07949.

\end{document}